\documentclass[pra,aps,twocolumn,nopacs,superscriptaddress,nofootinbib]{revtex4}
\usepackage{graphicx}
\usepackage{dcolumn}
\usepackage{bm}
\usepackage{amsmath}
\usepackage{epsfig}
\usepackage{color}

\def\ii{{\rm i}}  \def\ee{{\rm e}}
\def\rb{{\bf r}}  \def\gb{{\bf g}}  \def\pb{{\bf p}}  \def\kb{{\bf k}}
\def\db{{\bf d}}
\def\Rb{{\bf R}}  \def\Qb{{\bf Q}}    \def\Eb{{\bf E}}

\def\xx{\hat{\bf x}}  \def\yy{\hat{\bf y}}  \def\zz{\hat{\bf z}}
\def\kpar{{k_\parallel}}  \def\kbpar{{{\bf k}_\parallel}}
\def\kparb{{{\bf k}_\parallel}}
  \def\eb{{\bf e}}  \def\qb{{\bf q}}  \def\nt{\hat{\bf n}} \def\Sb{{\bf S}}

\begin{document}
\title{Lasing and Amplification from Two-Dimensional Atom Arrays}
\author{Vahagn~Mkhitaryan}
\thanks{These two authors contributed equally.}
\affiliation{ICFO-Institut de Ciencies Fotoniques, The Barcelona Institute of Science and Technology, 08860 Castelldefels (Barcelona), Spain}
\author{Lijun~Meng}
\thanks{These two authors contributed equally.}
\affiliation{ICFO-Institut de Ciencies Fotoniques, The Barcelona Institute of Science and Technology, 08860 Castelldefels (Barcelona), Spain}
\affiliation{State Key Laboratory of Modern Optical Instrumentation, Zhejiang University, Hangzhou 310027, China}
\author{Andrea~Marini}
\affiliation{ICFO-Institut de Ciencies Fotoniques, The Barcelona Institute of Science and Technology, 08860 Castelldefels (Barcelona), Spain}
\affiliation{Department of Physical and Chemical Sciences, University of L'Aquila, Via Vetoio, 67100 L'Aquila, Italy}
\author{F.~Javier~Garc\'{\i}a~de~Abajo}
\email[Corresponding author: ]{javier.garciadeabajo@nanophotonics.es}
\affiliation{ICFO-Institut de Ciencies Fotoniques, The Barcelona Institute of Science and Technology, 08860 Castelldefels (Barcelona), Spain}
\affiliation{ICREA-Instituci\'o Catalana de Recerca i Estudis Avan\c{c}ats, Passeig Llu\'{\i}s Companys 23, 08010 Barcelona, Spain}

\begin{abstract}
We explore the ability of two-dimensional periodic atom arrays to produce light amplification and generate laser emission when gain is introduced through external optical pumping. Specifically, we predict that lasing can take place for arbitrarily weak atomic scatterers assisted by cooperative interaction among atoms in a 2D lattice. We base this conclusion on analytical theory for three-level scatterers, which additionally reveals a rich interplay between lattice and atomic resonances. Our results provide a general background to understand light amplification and lasing in periodic atomic arrays, with promising applications in the generation, manipulation, and control of coherent photon states at the nanoscale.
\end{abstract}
\maketitle

\section{Introduction}

Periodic arrays of light scatterers have the ability to enhance the optical near-field intensity due to the accumulation of in-phase scattering wave components. This is neatly illustrated by an infinite linear array of point scatterers illuminated with a plane wave of momentum and electric field both perpendicular to the array direction \cite{R1907}: the field induced on any given scatterer by the rest of the array diverges as the series $1+1/2+1/3+\dots$ when the wavelength is equal to the period; this divergence prevents the induction of polarization on the scatterers, thus rendering the array invisible under these conditions. Such types of lattice-sum divergences lead to Wood's anomalies \cite{W1902,F1936}, extraordinary optical transmission \cite{ELG98}, complete optical reflection, and large near-field enhancement, which are phenomena generally describable in terms of lattice resonances \cite{paper090}. Interestingly, complete reflection is observed even in the limit of small scatterers at the cost of narrowing down the spectral features and lowering their tolerance to structural defects \cite{paper064,AB08}.

A good example of small scatterers is provided by lossless quantum emitters incorporating two nondegenerate electronic levels, which are well-known to offer an optical cross-section $3\lambda^2/2\pi$ for light of wavelength $\lambda$. For a properly designed focused light beam, an individual atom is predicted to produce complete reflection \cite{ZMS08}, while an experimental realization of this idea has achieved $>10\%$ extinction by an individual 2-level molecule \cite{RWL12}. A similar effect takes place in one-dimensional waveguides, where a single 2-level scatterer also leads to complete reflection \cite{HGA16}. Likewise, the ability of two-dimensional (2D) arrays of small scatterers to produce complete reflection \cite{paper064} has been theoretically illustrated by considering 2-level quantum emitters \cite{SWL17}, again relying on lattice resonances \cite{paper090}.

Quantum emitters with optical gain should produce an interesting interplay between lattice and atomic resonances. In a related context, lattice resonances can effectively act as optical cavities in 2D photonic crystals, leading to laser emission in the presence of gain media \cite{INC99,MMD99}. Additionally, defect modes in these types of structures exhibit high quality factors, so they can operate as light-wavelength-scale laser cavities \cite{INC99,NYI01,AEV06,WBS15}. Stimulated by theoretical studies of plasmon-based lasers \cite{BS03,KHR17}, 2D arrays of plasmonic scatterers have been shown to also serve as laser cavities \cite{VVG13,YHD15}. In these works, a gain medium is added to the dielectric or metal that forms the periodic structure of photonic crystals or periodic plasmonic arrays. A situation in which the scatterers act simultaneously as the gain medium constitutes a likely source of unexplored phenomena, for example in arrays of three-level scatterers controlled through an external optical pump.

\begin{figure}[b]
\includegraphics[width=0.5\textwidth]{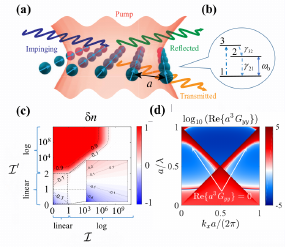}
\caption{{\bf (a)} Schematic view of a 2D square array (period $a$) of point scatterers with gain, pumped and probed with light plane waves of field amplitudes $\Eb^{\rm  pump}$ and $\Eb^{\rm probe}$, respectively. {\bf (b)} Energy diagram of a 3-level individual emitter defining the optical transition of frequency $\omega_0$ and the nonradiative damping rates $\gamma_{21}$ and $\gamma_{32}$. {\bf (c)} Population difference $\delta n$ of the system in (b) as a function of normalized pump and probe intensities (see main text). The inset is a zoom of the low-intensity region. {\bf (d)} Dipole lattice sum ${\rm Re}\{a^3 G_{yy}(\kbpar,\omega)\}$ for polarization along $y$ as a function of light wavelength $\lambda$ and parallel wave vector $\kbpar=\kpar\xx$ along the $x$ direction.}
\label{Fig1}
\end{figure}

Here, we investigate light amplification and lasing in 2D periodic arrays of externally-pumped 3-level atoms. A varied phenomenology is revealed, resulting from the interplay between lattice resonances and optical gain. In particular, laser emission is predicted to take place for atoms with arbitrarily weak transition strength, assisted by cooperative interaction through diverging lattice resonances. We reach these conclusions by formulating an analytical model in which the atoms are described through their polarizability obtained from a density-matrix formalism including gain, while the array periodicity enters through dipole-dipole lattice sums. Our results hold great potential for the design of atom-based optical-gain devices and novel sources of coherent radiation.

\section{Theoretical model}

For simplicity, we consider 3-level identical atoms [electronic energies $\hbar\varepsilon_1<\hbar\varepsilon_2<\hbar\varepsilon_3$, see Fig.\ \ref{Fig1}(b)] under resonant pump illumination at frequency $\omega'=\varepsilon_3-\varepsilon_1$. Following pumping from level 1 to 3, we assume the system to rapidly decay nonradiatively from 3 to 2. We are interested in the subsequent radiative decay from 2 to the ground state 1, which affects the scattering of light near the resonance frequency $\omega_0=\varepsilon_2-\varepsilon_1$, and ultimately gives rise to lasing (see below).

The internal temporal dynamics of the atoms in the array is governed by the Hamiltonian ${\cal H}(t)={\cal H}^{\rm at}+{\cal H}^{\rm rad}+{\cal H}^{\rm at-rad}+{\cal H}^{\rm ext}(t)$, where ${\cal H}^{\rm at}=\hbar\sum_{li}\varepsilon_i|li\rangle\langle li|$ and ${\cal H}^{\rm rad}=\hbar\sum_n\omega_n a_n^\dagger a_n$ describe the free atoms (electronic states $|li\rangle$ with $i=1-3$ for each of the atoms $l$ at positions $\rb_l$) and radiation (photon modes $n$ with creation and annihilation operators $a_n^\dagger$ and $a_n$), the term ${\cal H}^{\rm at-rad}=\sum_{nlii'}(g_{nlii'}^*a_n^\dagger+g_{nlii'} a_n)(\sigma_{lii'}^\dagger+\sigma_{lii'})$ accounts for light-atom interaction (coupling coefficients $g_{nlii'}$), ${\cal H}^{\rm ext}(t)=-\sum_{lii'}\db_{ii'}\cdot\Eb_l^{\rm ext}(t)(\sigma_{lii'}^\dagger+\sigma_{lii'})$ represents the interaction with the external field $\Eb_l^{\rm ext}(t)=\Eb^{\rm pump}(\rb_l)\ee^{-\ii\omega't}+\Eb_l^{\rm loc}\ee^{-\ii\omega t}+{\rm c.c.}$ [pump and local probe at frequencies $\omega'$ and $\omega$; see below for the connection between $\Eb_l^{\rm loc}$ and the probe field $\Eb^{\rm probe}(\rb_l)$], and we have defined atomic-transition operators $\sigma_{lii'}=|li\rangle\langle li'|$ and their corresponding dipole elements $\db_{ii'}=-e\langle li|\rb-\rb_l|li'\rangle$ (independent of $l$).

We treat the external field semi-classically and assume that the emitted photons are excited into coherent states \cite{SZ97,BS03,S10_2} (see Appendix for more details). This approximation allows us to factorize the density matrix of the entire system as the product of radiation and atomic subsystems $\rho=\rho^{\rm rad}\otimes\Pi_l\rho_l^{\rm at}$, substitute the photon operators by their complex-number expectation values, and write a self-contained equation of motion for each atom $l$ as $\dot{\rho}_l^{\rm at}=(\ii/\hbar)[\rho_l^{\rm at},{\cal H}(t)] + {\cal L}[\rho_l^{\rm at}]$, where the Lindblad term ${\cal L}[\rho_l^{\rm at}]=\sum_{ii'}(\gamma_{ii'}/2)\left(2\sigma_{li'i}\rho_l^{\rm at}\sigma_{li'i}^\dagger-\sigma_{li'i}^\dagger\sigma_{li'i}\rho_l^{\rm at}-\rho_l^{\rm at}\sigma_{li'i}^\dagger\sigma_{li'i}\right)$ describes nonradiative $i\rightarrow i'$ transitions [in practice, we only include $3\rightarrow2$ and $2\rightarrow1$ at rates $\gamma_{32}$ and $\gamma_{21}$, see Fig.\ \ref{Fig1}(b)].

At this point, we assume a uniform pump acting with the same strength on all atoms and an incident probe plane wave having a wave vector component $\kbpar$ parallel to the array \cite{array2D_comment1}. This wave vector is inherited by the linearly induced dipoles $\pb_l$, where the dependence on in-plane atom position $\rb_l=(x_l,y_l,0)$ comes from both the spatial variation of the external field and the relative atomic arrangement. Following a well-established procedure \cite{paper090,paper182}, the component of the induced dipoles at the probe frequency $\omega$ reduces to $\pb_l=\pb\,\ee^{\ii\kbpar\cdot\rb_l-\ii\omega t}+{\rm c.c.}$, where $\pb=\alpha(\omega)\cdot\Eb_0^{\rm loc}$ is a position-independent dipole amplitude (evaluated from the local probe field acting on the atom at position $\rb_{l=0}=0$) and $\alpha(\omega)$ is the atomic polarizability tensor (see below). Additionally, $\Eb_0^{\rm loc}$ is the sum of the incident probe $\Eb^{\rm probe}(0)$ and the field induced by the rest of the atoms, which admits the self-consistent form \cite{paper090,paper182} $\Eb_0^{\rm loc}=\left[1-G(\kbpar,\omega)\!\cdot\alpha(\omega)\right]^{-1}\!\!\cdot\Eb^{\rm probe}(0)$, where $G(\kbpar,\omega)=\sum_{l\neq0}[\omega^2/c^2+\nabla_{\rb_l}\otimes\nabla_{\rb_l}]\ee^{\ii\kbpar\cdot\rb_l}/r_l$ is a lattice sum that describes the electromagnetic dipole-dipole interactions, excluding self interactions ($l=0$ term).

The atomic polarizability is affected by the pump through changes in the population difference $\delta n=\rho^{\rm at}_{22}-\rho^{\rm at}_{11}$. (Note that under the assumed conditions all atoms are equally pumped, so their populations are independent of $l$.) A detailed nonperturbative solution of the equations of motion for the component of frequency $\omega$ under the rotating-wave approximation and neglecting higher-order harmonics allows us to obtain the induced dipoles directly from the expectation values $\pb=\sum_{ii'}\db_{ii'}{\rm tr}\{\rho_{l=0}^{\rm at}(\sigma_{ii'}^\dagger+\sigma_{ii'})\}$, from which the atomic polarizability is found to be (see Appendix for a detailed derivation)
\begin{align}
\alpha^{-1}(\omega) = \left[ \frac{2\omega_0\delta n}{\hbar} \frac{\db_{12}\otimes\db_{12}}{(\omega+\ii\gamma_{21}/2)^2-\omega_0^2} \right]^{-1} \!-\frac{2\ii\omega^3}{3c^3},
\nonumber
\end{align}
where the last term originates in the imaginary part of the dipole self-interaction, while the real part of this term is effectively absorbed as a vacuum resonance-frequency shift \cite{FT02}. The population difference admits an involved analytical expression that is derived in the Appendix. It is however illustrative to consider the $\gamma_{32}\gg\gamma_{21}$ limit near resonant probe illumination conditions ($\delta=\omega-\omega_0\ll\omega_0$), which permits us to write
\[\delta n=\frac{-1+\mathcal{I}'}{1+8\mathcal{I}(1+3\mathcal{I}'\gamma_{21}/2\gamma_{32})/(1+4\delta^2/\gamma_{21}^2)+\mathcal{I}'}.\]
Incidentally, we retain a term $\propto\mathcal{I}\mathcal{I}'$ in the denominator that produces saturation of lasing (see below). Here, $\mathcal{I}'=\left|\Eb^{\rm pump}/E^{\rm pump}_{\rm thres}\right|^2$ and $\mathcal{I}=\left|\Eb^{\rm loc}/E^{\rm probe}_{\rm sat}\right|^2$ are the pump and local probe field intensities normalized to their respective threshold and saturation values $E^{\rm pump}_{\rm thres}=2\hbar\sqrt{\gamma_{21}\gamma_{32}}/d_{13}$ and $E^{\rm probe}_{\rm sat}=\hbar\gamma_{21}/d_{12}$, respectively. We plot $\delta n$ for $\omega=\omega_0$ in Fig.\ \ref{Fig1}(c), which shows that the full range $\delta n\in[-1,1]$ is reached. In what follows, we use $\delta n$ as an input parameter controlled by the combination of pump and probe intensities.

For concreteness, we consider a square array of period $a$, illuminated by $s$ polarized light with $\kbpar$ along one of the principal axes $\xx$, so that the optical electric field and the induced dipoles are both aligned along the remaining lattice axis $\yy$ [Fig.\ \ref{Fig1}(a)]. The response of the array is then captured by its specular-reflection and transmission coefficients, which can be written from the induced dipoles as \cite{paper090,paper182}
\begin{align}
r&=\ii S/[1/\alpha(\omega)-G_{yy}(\kbpar,\omega)], \label{rr}\\
t&=1+r,
\nonumber
\end{align}
where $S=2\pi\omega/c a^2\cos\theta$, and $\theta$ is the angle of incidence (see Appendix). These coefficients are dominated by the $\omega=\omega_0$ pole of $\alpha(\omega)$ and the lattice resonances of $G_{yy}(\kbpar,\omega)$, the real part of which is plotted in Fig.\ \ref{Fig1}(d).

\begin{figure}[t]
\includegraphics[width=0.5\textwidth]{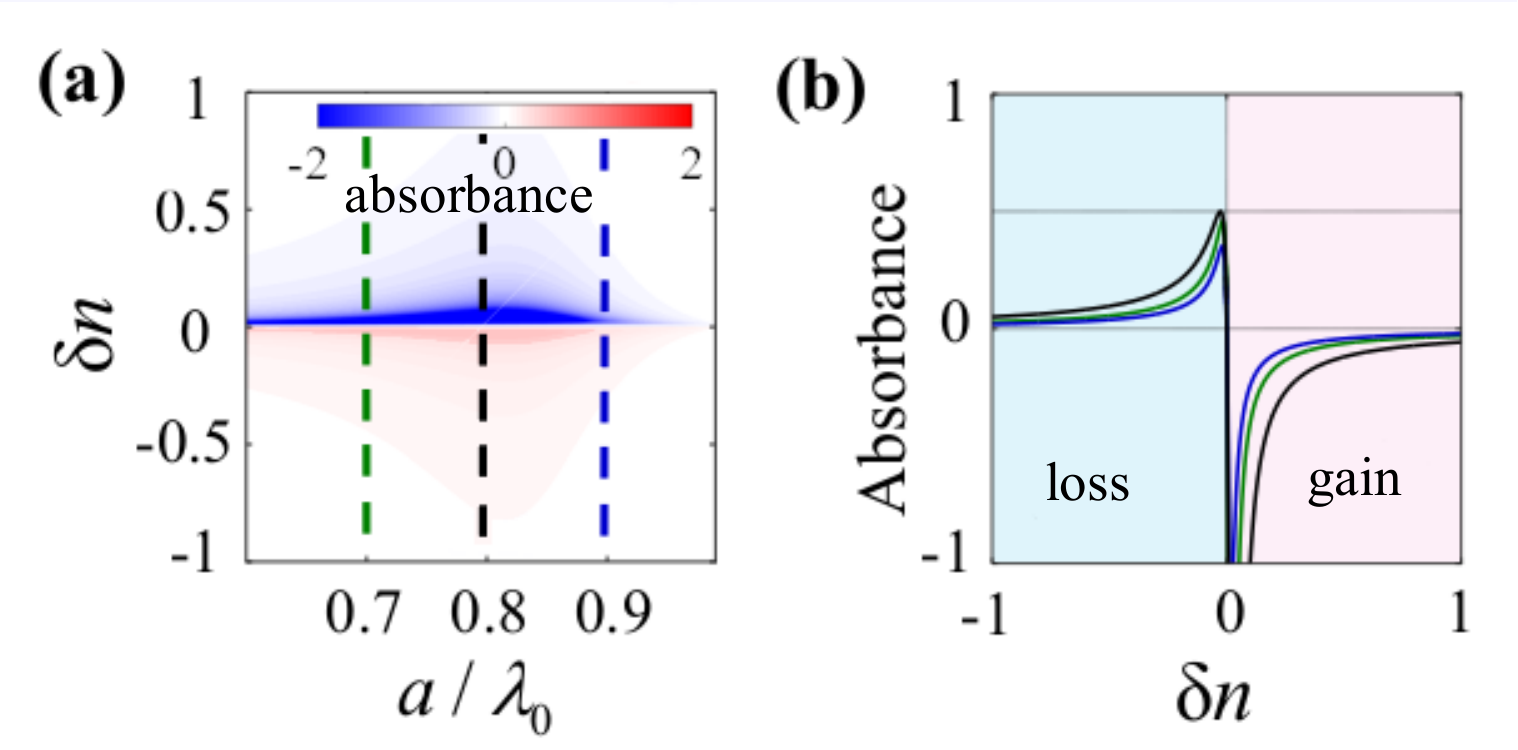}
\caption{{\bf (a)} Absorbance under normal-incidence resonant-wavelength ($\lambda=\lambda_0=2\pi c/\omega_0$) probe illumination conditions as a function of population difference $\delta n$ and period-to-wavelength ratio $a/\lambda_0$. {\bf (b)} Cuts through (a) along the indicated vertical dashed lines with the same color code. We take the ratio between nonradiative and radiative scatterer decay rates to be $\gamma_{21}/\gamma_0=0.01$.}
\label{Fig2}
\end{figure}

\section{Optical gain in 2D atom arrays}

We conclude from the above analysis that the external pump enables active tuning of the atomic polarizability $\alpha(\omega)$ by controlling $\delta n$. In particular, population inversion ($\delta n > 0$) leads to amplification of the probe, which, combined with gain from each of the atoms in the 2D array and collective excitations associated with lattice resonances, gives rise to peculiar features in the absorbance $1-|r|^2-|t|^2$. This is illustrated in Fig.\ \ref{Fig2} for resonant illumination (wavelength $\lambda_0 = 2\pi c/\omega_0$) under normal incidence as a function of $\delta n$ and $a/\lambda_0$. We assume a small nonradiative decay rate $\gamma_{21}=0.01\,\gamma_0$ compared with the natural radiative decay rate $\gamma_0=4\omega_0^3d_{12}^2/3\hbar c^3$. For moderate pumping ($-1<\delta n < 0$), the absorbance increases with $\delta n$ until it reaches the maximum theoretical limit of $0.5$ near $a/\lambda_0=0.8$ and $\delta n\lesssim0$ [Fig.\ \ref{Fig2}(b)]. For higher pumping, we have $0<\delta n<1$, leading to a peak of negative absorbance [Fig.\ \ref{Fig2}(b)], which is signalled by a minimum of $\left|1/\alpha-G_{yy}\right|$ [i.e., a lattice resonance, see Eq.\ (\ref{rr})].

\begin{figure*}[t]
\includegraphics[width=1\textwidth]{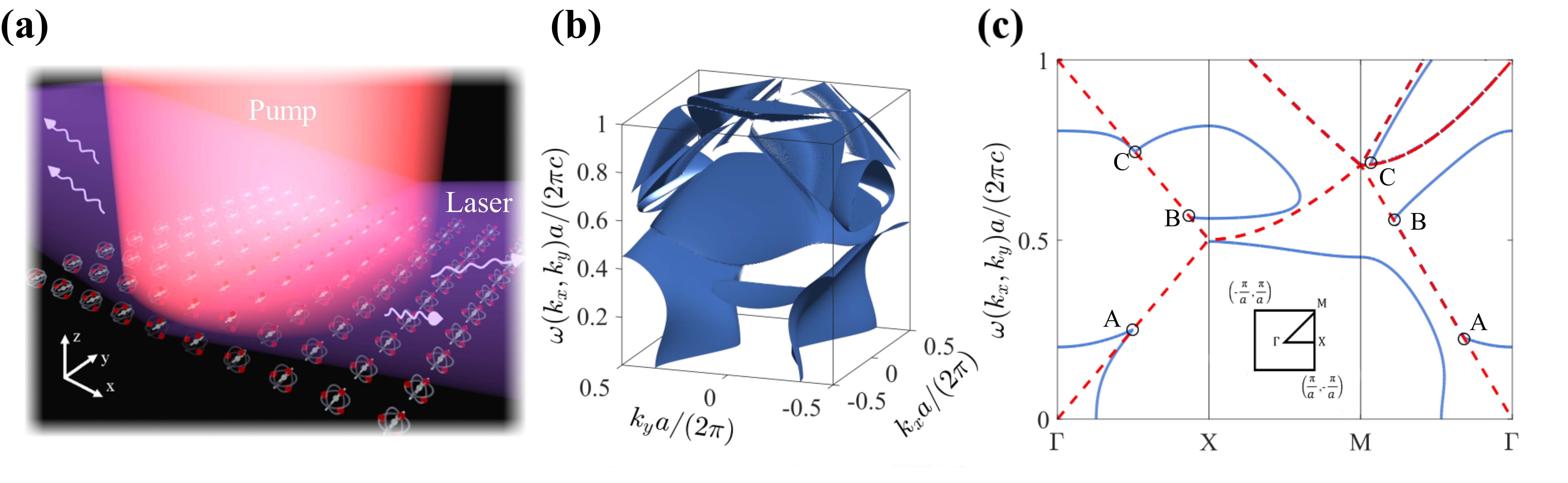}
\caption{{\bf (a)} Geometry considered for lasing in the 2D atomic array. {\bf (b)} Parallel-wave-vector-dependent frequency surfaces $\omega(\kbpar)$ corresponding to the lasing condition ${\rm Re}\{G_{yy}(\kbpar,\omega)\}=0$. {\bf (c)} Cuts of (b) along a characteristic excursion within the irreducible Brillouin zone (solid curves), along with the dispersion of the empty lattice (dashed curves). We indicate singular crossing points A-C (see main text).}
\label{Fig3}
\end{figure*}

\begin{figure}[t]
\includegraphics[width=0.4\textwidth]{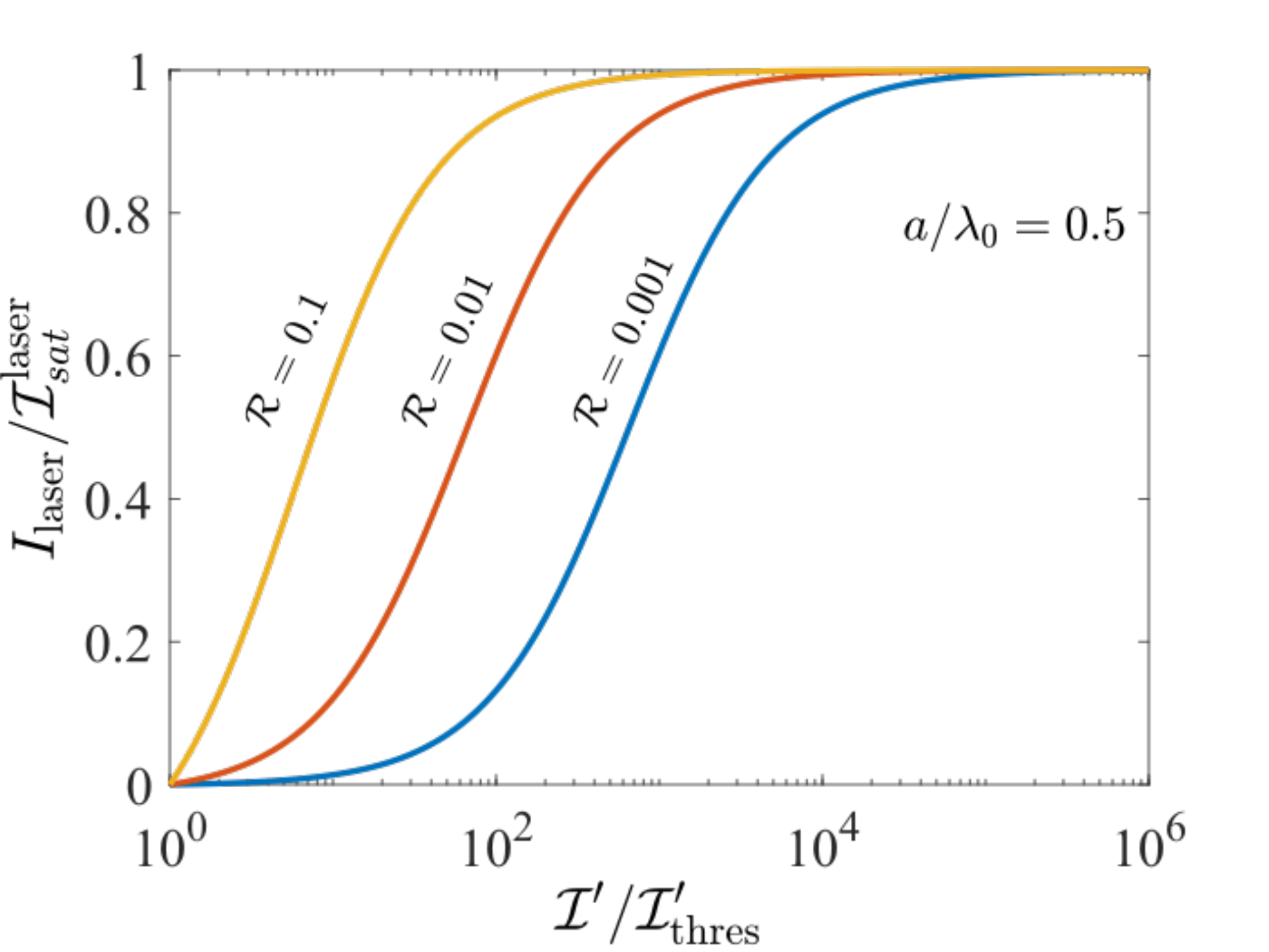}
\caption{Normalized lasing intensity as a function of normalized pump intensity for different values of the ratio $\mathcal{R}=\gamma_{21}/\gamma_{32}$ and fixed values of $a=\lambda_0/2$ and $\gamma_0=0.01\gamma_{21}$.}
\label{Fig4}
\end{figure}

\section{Directional lasing}

Lattice resonances are signaled by maxima of the reflectivity [Eq.\ (\ref{rr})], which in the presence of gain may become divergent, provided radiative losses are compensated. This leads to sustained lasing. Assuming a square lattice of three-level atoms with their $\omega_0$ transition dipoles all oriented along the $y$ direction under normal-incidence pumping [Fig.\ \ref{Fig3}(a)], we find resonances signaled by the condition $\alpha(\omega) G_{yy}(\kbpar,\omega)=1$ according to Eq.\ (\ref{rr}). Additionally, energy conservation imposes a balance between pump excitation, inelastic absorption, and laser emission: $I^{\rm pump} \mathcal{A}' - I^{\rm laser} - I^{\rm nr} = 0$, where $\mathcal{A}'$ is the absorbance at the pump frequency, $I^{\rm nr} = \rho_{22}\gamma_{21}\hbar\omega_0/a^2+\rho_{33}\gamma_{32}\hbar(\omega'_0-\omega_0)/a^2$ represents the power per unit area associated with nonradiative processes, and $I^{\rm laser}$ is the emission intensity produced by the self-consistently induced dipoles $\pb$. Following well-established methods (see Appendix), we find that these two requirements are simultaneously fulfilled when the emission occurs at frequency $\omega=\sqrt{\omega_0^2+\gamma_{21}^2/4}$, under the condition ${\rm Re}\{G_{yy}(\kbpar,\omega)\}=0$. The stable population difference is then fixed by
\begin{align}
\frac{1}{\delta n}=\frac{3\pi c^3\gamma_0}{\gamma_{21}\omega_0^2\omega a^2}\sum_\gb {\rm Re}\left\{\frac{\omega^2/c^2-(k_y+g_y)^2}{\sqrt{\omega^2/c^2-|\kbpar+\gb|^2}}\right\}, \label{deltan_1}
\end{align}
where the sum runs over reciprocal lattice vectors $\gb$. Moreover, the laser intensity reduces to
\begin{align}
I^{\rm laser} = \frac{\hbar\omega_0\gamma_{21}}{2a^2}\;
\frac{(1-\mathcal{R})\mathcal{I}'-1
-\left[(1+2\mathcal{R})\mathcal{I}'+1\right]\,\delta n}{1+(3/2)\mathcal{R}\mathcal{I}'}
\nonumber
\end{align}
with $\mathcal{R}=\gamma_{21}/\gamma_{32}$ \cite{array2D}, which describes a typical lasing behavior as a function of pump intensity (see Fig. \ref{Fig4}).

\begin{figure}[t]
\includegraphics[width=0.45\textwidth]{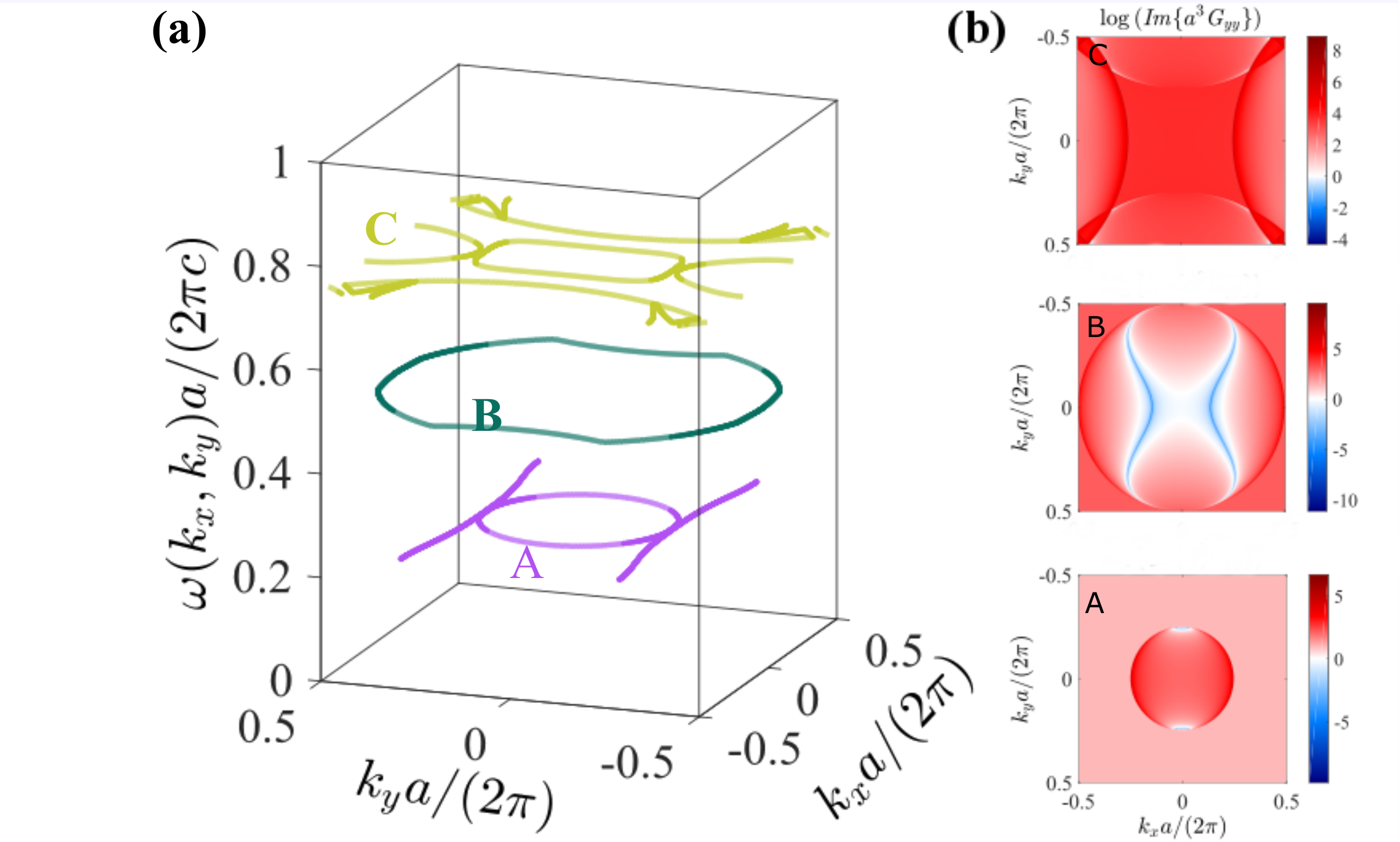}
\caption{{\bf (a)} Evolution of the singular points A-C [see Fig.\ \ref{Fig3}(c)] in $\kbpar$-$\omega$ space. {\bf (b)} Maps of ${\rm Im}\{G_{yy}\}$ at fixed $\omega$ planes near the A-C contours.}
\label{Fig5}
\end{figure}

The low-frequency ${\rm Re}\{G_{yy}(\kbpar,\omega)\}=0$ surfaces for this system are plotted in Fig.\ \ref{Fig3}(b). We remark that each surface point corresponds to a different direction of emission determined by $\kbpar$. Loss compensation and lasing from the 2D array then require a specific value of the atom dipole $\pb$, which is only possible when optical pumping $\mathcal{I}'$ overcomes the threshold $\mathcal{I}'_{\rm thres}=(1+\delta n)/[1-\mathcal{R}-(1+2\mathcal{R})\,\delta n]$ (see Appendix). Obviously, this expression can only be satisfied if $\delta n<(1-\mathcal{R})/(1+2\mathcal{R})\equiv\delta n_\mathcal{R}$, which imposes a minimum transition strength $d_{12}^2\propto\gamma_0$ according to Eq.\ (\ref{deltan_1}).

\section{Lasing from arbitrarily weak atoms}

The atomic transition strength controls the population difference $\delta n$ through $\gamma_0$ [see Eq.\ (\ref{deltan_1})]. Now, even weak atoms (i.e., having small $\gamma_0$) can sustain lasing when the right-hand side (rhs) of that equation is compensated by a lattice resonance ($|\kbpar+\gb|\approx\omega/c$). Lattice resonances for different $\gb$'s are plotted in Fig.\ \ref{Fig3}(c) (dashed curves), where we identify crossings with the lasing condition ${\rm Re}\{G_{yy}\}=0$ (solid curves), corresponding to three different 1D contours A-C, also represented in Fig.\ \ref{Fig5}(a) as a function of $\kbpar$. Importantly, these are real divergences of the rhs of Eq.\ (\ref{deltan_1}) directly inherited from ${\rm Im}\{G_{yy}\}$ [see Fig.\ \ref{Fig5}(b)]. We thus conclude that lasing can take place near these resonances regardless of how small $\gamma_0$ is.

\section{Conclusion}

Our study demonstrates that the interplay between lattice resonances and gain from externally pumped atoms arranged in 2D periodic arrays not only produces resonant amplification of scattered light, but also lasing emission even for arbitrarily weak atoms. These results have general applicability to 3-level atom systems, including ultracold trapped atom arrays, atomic clusters, and macroscopic meta-atoms. Also, they can be readily extended to other atomic electronic structures, or even to particles containing a large number $N$ of optically pumped atoms, for which the effective atomic radiative decay rate $\gamma_0$ is simply multiplied by a factor $N$. We envision a generalization of the present results to more involved lattices for the development of 2D lasing metasurfaces with engineered polarization, intensity, and phase emission patterns.

\begin{widetext}

\appendix

\begin{figure}
\begin{centering}
\includegraphics[width=0.3\textwidth]{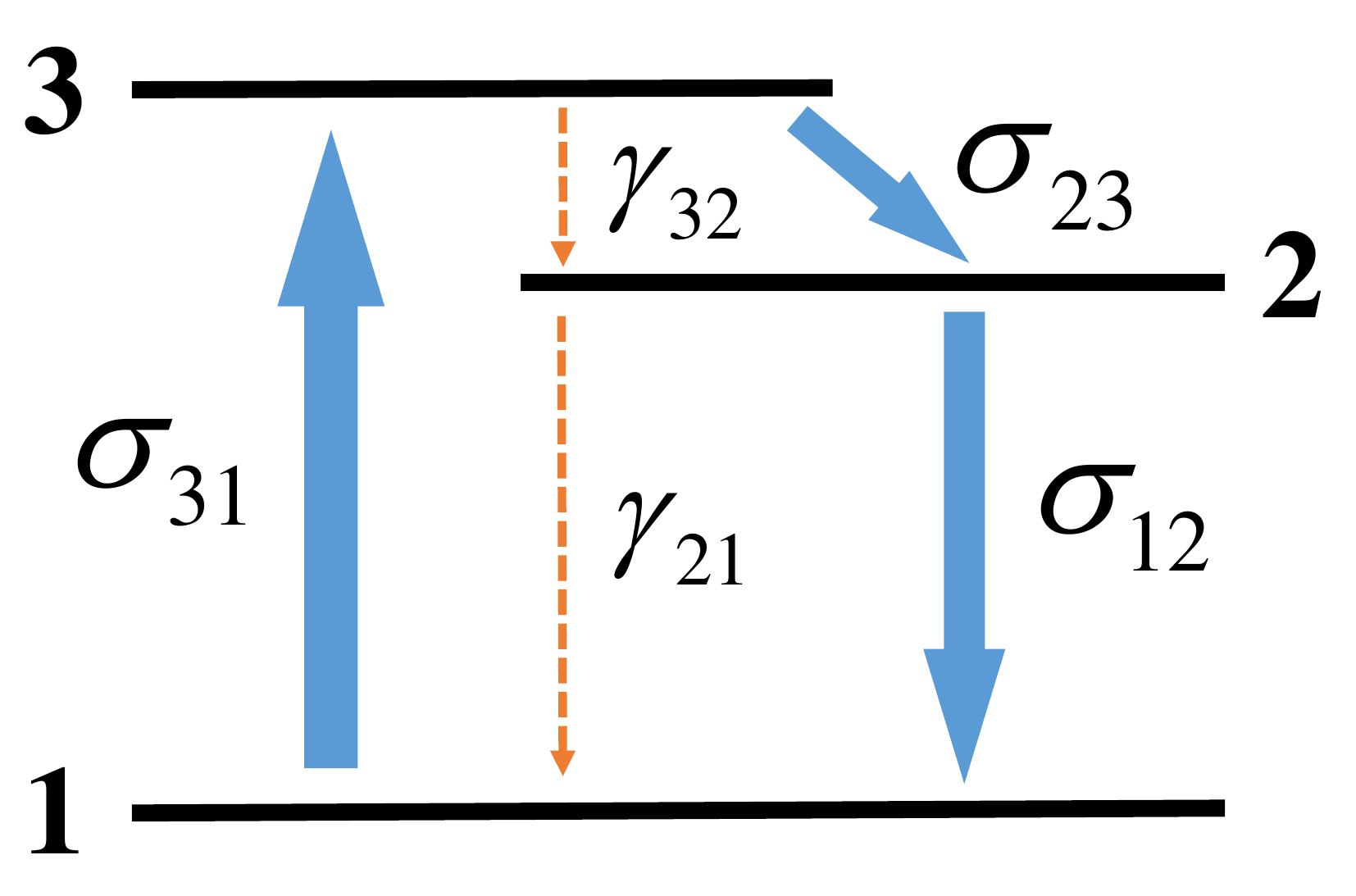}	
\par\end{centering}
\caption{Sketch of the 3-level system considered in our calculations.}
\label{FigS1}
\end{figure}

\section{Quantum dynamics in arrays of three-level emitters}

\subsection{Density-matrix description}

We consider a system of identical 3-level optical emitters (labeled by $l$) coupled to a bath of photons (boson modes labeled by $n$). The temporal dynamics of this system can be generally described through the time-dependent Hamiltonian
\begin{align}
{\cal H} = \hbar\sum_{li} \varepsilon_i|li\rangle\langle li| + \hbar\sum_n \omega_n a_n^\dagger a_n + \sum_{nlii'} \left( g_{nlii'}^*a_n^\dagger  + g_{nlii'}a_n \right) \left( \sigma_{lii'}^\dagger  + \sigma_{lii'} \right) - \sum_{lii'} \db_{ii'}\cdot\Eb^{\rm ext}_l(t) \left( \sigma_{lii'}^\dagger  + \sigma_{lii'} \right),
\nonumber
\end{align}
where the index $i=1-3$ runs over the emitter levels of energies $\hbar\varepsilon_i$; the operators $\sigma_{lii'}=|li\rangle\langle li'|$ describe $i'\rightarrow i$ atomic transitions of the emitter $l$ at the position $\rb_l$; the operators $a_n^\dagger$ and $a_n$ account for the creation and annihilation of photons with energy $\hbar\omega_n$; the complex coupling coefficients $g_{nlii'}$ connect photons in mode $n$ and the levels $i$ and $i'$ in emitter $l$; the compex vectors $\db_{ii'}$ are the corresponding emitter dipole moments; and $\Eb^{\rm ext}_l(t) = \Eb^{\rm pump}(\rb_l)\ee^{-\ii{\omega'} t} + \Eb^{\rm probe}(\rb_l)\ee^{-\ii\omega t} + {\rm c.c.}$ is the time-dependent external field at the position $\rb_l$ given by the superposition of pump and probe fields with amplitudes $\Eb^{\rm pump}(\rb_l)$ and $\Eb^{\rm probe}(\rb_l)$, oscillating with frequencies $\omega'$ and $\omega$, respectively. Incidentally, $\sigma_{lii'}^\dagger=\sigma_{li'i}$, so in the double sums over $ii'$ throughout this document we only consider terms $i>i'$ in order to avoid counting these transitions twice. The temporal evolution of the total density matrix $\rho(t)$ is governed by the equation of motion
\begin{align}
\dot{\rho} = - \frac\ii{\hbar}[{\cal H}(t),\rho] + {\cal L}[\rho], \label{DensityMatrixEq}
\end{align}
in which the commutator $[{\cal H}(t),\rho]$ describes the coherent temporal evolution of the system, complemented by nonradiative incoherent transitions introduced through the Lindblad operator 
\begin{align}
{\cal L}[\rho] &= \sum_{lii'} \frac{\gamma_{ii'}}{2}\left(2\sigma_{lii'}\rho\sigma_{lii'}^\dagger  - \sigma_{lii'}^\dagger \sigma_{lii'}\rho - \rho\sigma_{lii'}^\dagger \sigma_{lii'} \right) \nonumber \\
&+ \sum_n \frac{\Gamma_n}{2}\left(2a_n\rho a_n^\dagger  - a_n^\dagger a_n\rho - \rho a_n^\dagger a_n \right),
\nonumber
\end{align}
with inelastic rates $\gamma_{ii'}$ for the atoms (see sketch in Fig.\ \ref{FigS1}) and $\Gamma_n$ for the photon mode $n$.

At this point, we approximate the density matrix by the tensor product $\rho = \rho^{\rm rad}\otimes\rho^{\rm at}$, where $\rho^{\rm at} = \Pi_l\rho_l^{\rm at}$, $\rho^{\rm rad} = \Pi_n|\alpha_n\rangle\langle\alpha_n|$, and we assume each photon state $n$ to be in a coherent state $|\alpha_n\rangle$ with amplitude $\alpha_n ={\rm Tr}\left\{a_n\rho\right\}$. Using $\left[a_n,a_{n'}^\dagger\right] = \delta_{nn'}$, along with the property $a_n|\alpha_n\rangle = \alpha_n|\alpha_n\rangle$ of coherent states (leading to $a_n\rho = \alpha_n\rho$ and $\rho a_n^\dagger  = \alpha_n^*\rho$), we find from Eq.\ (\ref{DensityMatrixEq}) that the coherent-state amplitude satisfies the equation \cite{SZ97,LSB03,S10_2}
\begin{align}
\dot{\alpha}_n = \frac{d}{dt}{\rm Tr} \left\{ a_n\rho \right\} = {\rm Tr} \left\{ a_n\dot{\rho} \right\} = -\ii\left(\omega_n-\ii\frac{\Gamma_n}{2}\right)\,\alpha_n - \frac\ii{\hbar}\sum_{lii'} g_{nlii'}^* {\rm Tr} \left\{ (\sigma_{lii'}^\dagger  + \sigma_{lii'})\rho^{\rm at} \right\}. \label{dpcmeq}
\end{align}  
Having assumed coherent states for the photons, the Hamiltonian of the system reduces to
\begin{align}
{\cal H} = \hbar\sum_n \omega_n |\alpha_n|^2 + \sum_l {\cal H}_l,
\nonumber
\end{align}
where
\begin{align}
{\cal H}_l = \hbar\sum_{i} \varepsilon_i|li\rangle\langle li| + \sum_{ii'} \left( g_{lii'}^* + g_{lii'} \right) \left( \sigma_{lii'}^\dagger  + \sigma_{ii'} \right) - \sum_{lii'} \db_{ii'}\cdot\Eb^{\rm ext}_l(t) \left(\sigma_{lii'}^\dagger  + \sigma_{lii'} \right),
\nonumber
\end{align}
and we introduce the new coupling parameters
\begin{align}
g_{lii'} = \sum_n g_{nlii'}\alpha_n.
\label{gliibis}
\end{align}

We assume that every emitter can be described as the 3-level system sketched in Fig.\ \ref{FigS1}. Then, the equation of motion of a single emitter is explicitly given by
\begin{align}
\dot{\rho}_l = - \frac\ii{\hbar}[{\cal H}_l,\rho_l] + \frac{\gamma_{21}}{2}\left(2 \sigma_{l12} \rho_l \sigma_{l12}^\dagger  - \sigma_{l12}^\dagger  \sigma_{l12} \rho_l - \rho_l \sigma_{l12}^\dagger  \sigma_{l12} \right) + \frac{\gamma_{32}}{2}\left(2 \sigma_{l23} \rho_l \sigma_{l23}^\dagger  - \sigma_{l23}^\dagger  \sigma_{l23} \rho_l - \rho_l \sigma_{l23}^\dagger  \sigma_{l23} \right),
\nonumber
\end{align}
where we drop the at superscript from $\rho_l^{\rm at}$ for convenience. In what follows, we also omit the index $l$, which is implicitly understood in the following expressions. Note that we assume the damping rates $\gamma_{21}$ and $\gamma_{32}$, the energy levels $\hbar\varepsilon_i$, and the transition dipoles $\db_{12}$ and $\db_{13}$ to be the same for all emitters (i.e., independent of $l$). Damping rates and transition dipoles other than these ones are disregarded. Additionally, radiative damping from level 3 is ignored, so $g_{12}$ is the only nonzero radiative coupling coefficient. Now, we express the density matrix in the state representation $\rho = \sum_{ii'}\rho_{ii'}|i\rangle\langle i'|$, from which the property $\rho^\dagger=\rho$ is found to lead to the condition $\rho_{ii'}=\rho^*_{i'i}$, so we only need to work out the elements with $i\le i'$. Explicitly, the density-matrix equations of motion read
\begin{align}
\dot{\rho}_{11} & = \frac{2}{\hbar}\left[\db_{12}\cdot\Eb^{\rm probe}(t) - \left(g_{12}^*+g_{12}\right) \right] {\rm Im}\{\rho_{12}\} + \frac{2}{\hbar} \db_{13}\cdot\Eb^{\rm pump}(t) {\rm Im}\{\rho_{13}\} + \gamma_{21}\rho_{22}, \nonumber\\
\dot{\rho}_{22} & = - \frac{2}{\hbar}\left[\db_{12}\cdot\Eb^{\rm probe}(t) - \left(g_{12}^*+g_{12}\right) \right] {\rm Im}\{\rho_{12}\} + \gamma_{32}\rho_{33} - \gamma_{21}\rho_{22}, \nonumber\\
\dot{\rho}_{33} & = - \frac{2}{\hbar} \db_{13}\cdot\Eb^{\rm pump}(t) {\rm Im}\{\rho_{13}\} - \gamma_{32}\rho_{33}, \nonumber\\
\dot{\rho}_{12} & = \ii\omega_0\rho_{12} + \frac\ii{\hbar}\left[ \db_{12}\cdot\Eb^{\rm probe}(t) - \left(g_{12}^*+g_{12}\right) \right] (\rho_{22}-\rho_{11}) - \frac{\gamma_{21}}{2}\rho_{12}, \nonumber\\
\dot{\rho}_{13} & = \ii\omega'_0\rho_{13} + \frac\ii{\hbar}\db_{13}\cdot\Eb^{\rm pump}(t) (\rho_{33}-\rho_{11}) - \frac{\gamma_{32}}{2}\rho_{13}, \nonumber
\end{align}
where we have defined
\begin{align}
\omega_0&=\varepsilon_2-\varepsilon_1, \nonumber\\
\omega'_0&=\varepsilon_3-\varepsilon_1, \nonumber
\end{align}
and we have assumed the optical pump to be nearly resonant with the $1\rightarrow 3$ transition ($\omega'\simeq\omega'_0$) and the probe field to be nearly resonant with the $1\rightarrow 2$ transition ($\omega\simeq\omega_0$). Separating real and imaginary parts of the coherences $\rho_{12} = \rho_{12}^{\rm R} + \ii\rho_{12}^{\rm I}$ and $\rho_{13} = \rho_{13}^{\rm R} + \ii\rho_{13}^{\rm I}$, one finds
\begin{align}
\dot{\rho}_{11} & = \frac{2}{\hbar}\left[ \db_{12}\cdot\Eb^{\rm probe}(t) - \left(g_{12}^*+g_{12}\right) \right] \rho_{12}^{\rm I} + \frac{2}{\hbar} \db_{13}\cdot\Eb^{\rm pump}(t) \rho_{13}^{\rm I} + \gamma_{21}\rho_{22}, \nonumber\\
\dot{\rho}_{22} & = - \frac{2}{\hbar}\left[ \db_{12}\cdot\Eb^{\rm probe}(t) - \left(g_{12}^*+g_{12}\right) \right] \rho_{12}^{\rm I} + \gamma_{32}\rho_{33} - \gamma_{21}\rho_{22}, \nonumber\\
\dot{\rho}_{33} & = - \frac{2}{\hbar} \db_{13}\cdot\Eb^{\rm pump}(t) \rho_{13}^{\rm I} - \gamma_{32}\rho_{33}, \nonumber\\
\dot{\rho}_{12}^{\rm R} & = - \omega_0\rho_{12}^{\rm I} - \frac{\gamma_{21}}{2}\rho_{12}^{\rm R}, \nonumber\\
\dot{\rho}_{12}^{\rm I} & = \omega_0\rho_{12}^{\rm R} + \frac{1}{\hbar}\left[ \db_{12}\cdot\Eb^{\rm probe}(t) - \left(g_{12}^*+g_{12}\right) \right] (\rho_{22}-\rho_{11}) - \frac{\gamma_{21}}{2}\rho_{12}^{\rm I}, \nonumber\\
\dot{\rho}_{13}^{\rm R} & = -\omega'_0\rho_{13}^{\rm I} - \frac{\gamma_{32}}{2}\rho_{13}^{\rm R}, \nonumber\\
\dot{\rho}_{13}^{\rm I} & = \omega'_0\rho_{13}^{\rm R} + \frac{1}{\hbar} \db_{13}\cdot\Eb^{\rm pump}(t) (\rho_{33}-\rho_{11}) - \frac{\gamma_{32}}{2}\rho_{13}^{\rm I}. \nonumber
\end{align}
Then, eliminating $\rho_{12}^{\rm I}$ and $\rho_{13}^{\rm I}$, we obtain
\begin{subequations}
\begin{align}
& \dot{\rho}_{11} = -\frac{2}{\hbar\omega_0}\left[ \db_{12}\cdot\Eb^{\rm probe}(t) - \left(g_{12}^*+g_{12}\right) \right] \left( \dot{\rho}_{12}^{\rm R} + \frac{\gamma_{21}}{2}\rho_{12}^{\rm R} \right) - \frac{2}{\hbar\omega'_0}\db_{13}\cdot\Eb^{\rm pump}(t)  \left( \dot{\rho}_{13}^{\rm R} + \frac{\gamma_{32}}{2}\rho_{13}^{\rm R} \right) + \gamma_{21}\rho_{22},  \\
& \dot{\rho}_{22} = \frac{2}{\hbar\omega_0}\left[ \db_{12}\cdot\Eb^{\rm probe}(t) - \left(g_{12}^*+g_{12}\right) \right] \left( \dot{\rho}_{12}^{\rm R} + \frac{\gamma_{21}}{2}\rho_{12}^{\rm R} \right) + \gamma_{32}\rho_{33} - \gamma_{21}\rho_{22},  \\
& \dot{\rho}_{33} = \frac{2}{\hbar\omega'_0} \db_{13}\cdot\Eb^{\rm pump}(t) \left( \dot{\rho}_{13}^{\rm R} + \frac{\gamma_{32}}{2}\rho_{13}^{\rm R} \right) - \gamma_{32}\rho_{33},  \\
& \ddot{\rho}_{12}^{\rm R} + \gamma_{21}\dot{\rho}_{12}^{\rm R} + \left(\omega_0^2+\frac{\gamma_{21}^2}{4}\right)\rho_{12}^{\rm R} = - \frac{\omega_0}{\hbar}(\rho_{22}-\rho_{11})\left[ \db_{12}\cdot\Eb^{\rm probe}(t) - \left(g_{12}^*+g_{12}\right) \right] ,  \\
& \ddot{\rho}_{13}^{\rm R} + \gamma_{32}\dot{\rho}_{13}^{\rm R} + \left({\omega'}_0^2+\frac{\gamma_{32}^2}{4}\right)\rho_{13}^{\rm R} = - \frac{\omega'_0}{\hbar}(\rho_{33}-\rho_{11})\,\db_{13}\cdot\Eb^{\rm pump}(t),  \\
& \rho_{12}^{\rm I} = - \frac{1}{\omega_0} \left( \dot{\rho}_{12}^{\rm R} + \frac{\gamma_{21}}{2}\rho_{12}^{\rm R}\right),  \\
& \rho_{13}^{\rm I} = - \frac{1}{\omega'_0} \left( \dot{\rho}_{13}^{\rm R} + \frac{\gamma_{32}}{2}\rho_{13}^{\rm R}\right). 
\end{align}
\label{dmeom}
\end{subequations}
It is straightforward to verify the condition $\sum_{l}\dot{\rho}_{ll}=0$, confirming that the total population $\sum_{l}\rho_{ll}=1$ is conserved.

\subsection{Continuous-wave solution, atomic polarizability, and coupled-dipoles equations}
 
Here, we derive the steady-state operation conditions for continuous-wave pump and probe at frequencies $\omega'$ and $\omega$, respectively. For the sake of compactness, we introduce the new variables $\eta$ and $\eta'$, accounting for the probe and pump electric field amplitudes and implicitly defined by
\begin{subequations}
\begin{align}
& \frac{1}{\hbar} \left[ \db_{12}\cdot \Eb^{\rm probe}(t) - (g_{12}^*+g_{12}) \right] =
 \frac{1}{\hbar} \db_{12}\cdot \Eb^{\rm loc}(t) = \eta  \ee^{-\ii\omega t}  + {\rm c.c.}, \label{etaeta} \\
& \frac{1}{\hbar}        \db_{13}\cdot \Eb^{\rm pump}(t)                              = \eta' \ee^{-\ii\omega' t} + {\rm c.c.},
\label{etapetap}
\end{align}
\end{subequations}
where the $g_{12}$ terms, representing the induced field (see below), have been absorbed into the local field $\Eb^{\rm loc}$.
Inserting these expressions into the density-matrix equations (\ref{dmeom}), anticipating the steady-state time dependences $\dot{\rho}_{11}=\dot{\rho}_{22}=\dot{\rho}_{33}=0$,
\begin{subequations}
\begin{align}
\rho_{12}^{\rm R} & = \xi  \ee^{-\ii\omega t } + {\rm c.c.}, \nonumber\\
\rho_{13}^{\rm R} & = \xi' \ee^{-\ii\omega' t} + {\rm c.c.}, \nonumber 
\end{align}
\end{subequations}
and adopting the rotating-wave approximation (RWA), we find
\begin{subequations}
\begin{align}
\rho_{11} & = \frac{\displaystyle \left( 1 + \gamma_{32}^{-1}\Delta' \right)\left( 1 + \gamma_{21}^{-1}\Delta \right)}{\displaystyle 1 + 2\gamma_{21}^{-1}\Delta + \left(\gamma_{21}^{-1}+2\gamma_{32}^{-1}\right)\Delta'+ 3\gamma_{21}^{-1}\gamma_{32}^{-1}\Delta\Delta' }, \label{eqrho11}\\
\rho_{22} & = \frac{\displaystyle \gamma_{21}^{-1}(\Delta+\Delta') + \gamma_{21}^{-1}\gamma_{32}^{-1}\Delta\Delta'}{\displaystyle 1 + 2\gamma_{21}^{-1}\Delta + \left(\gamma_{21}^{-1}+2\gamma_{32}^{-1}\right)\Delta'+ 3\gamma_{21}^{-1}\gamma_{32}^{-1}\Delta\Delta' }, \label{eqrho22}\\
\rho_{33} & = \frac{\displaystyle \gamma_{32}^{-1}\Delta' \left(1+\gamma_{21}^{-1}\Delta\right)}{\displaystyle 1 + 2\gamma_{21}^{-1}\Delta + \left(\gamma_{21}^{-1}+2\gamma_{32}^{-1}\right)\Delta'+ 3\gamma_{21}^{-1}\gamma_{32}^{-1}\Delta\Delta' }, \label{eqrho33}\\
\xi       & = \frac{\displaystyle \omega_0\eta}{\displaystyle (\omega+\ii\gamma_{21}/2)^2-\omega_0^2}
\left(\rho_{22}-\rho_{11}\right), \\
\xi'      & = \frac{\displaystyle \omega'_0\eta'}{\displaystyle (\omega'+\ii\gamma_{32}/2)^2-{\omega'}_0^2} \left(\rho_{33}-\rho_{11}\right),
\end{align}
\label{finalrho}
\end{subequations}
where we use
\begin{subequations}
\begin{align}
\Delta & = 4|\eta|^2{\rm Im} \left\{ \frac{-1}{(\omega  + \ii \gamma_{21}/2 ) - \omega_0^2/(\omega  + \ii \gamma_{21}/2 )} \right\} = \frac{\displaystyle 2\gamma_{21}|\eta|^2\left( \omega^2 + \omega_0^2 + \gamma_{21}^2/4 \right)}{ \displaystyle \left( \omega^2 - \omega_0^2 - \gamma_{21}^2/4\right)^2 + \gamma_{21}^2\omega^2}, \label{Delta} \\
\Delta' & = 4|\eta'|^2{\rm Im} \left\{ \frac{-1}{(\omega' + \ii \gamma_{32}/2 ) - {\omega'}_0^2/(\omega' + \ii \gamma_{32}/2 )} \right\} = \frac{\displaystyle 2\gamma_{32}|\eta'|^2\left( {\omega'}^2 + {\omega'_0}^2 + \gamma_{32}^2/4 \right)}{ \displaystyle \left( {\omega'}^2 - {\omega'}_0^2 - \gamma_{32}^2/4\right)^2 + \gamma_{32}^2{\omega'}^2}. \label{Deltap}
\end{align}
\end{subequations}
Now, the electric dipole moment induced in the atom is given by $\pb = \sum_{ii'}\db_{ii'}\,{\rm Tr} \left\{\left( \sigma_{lii'}^\dagger  + \sigma_{lii'} \right)\rho\right\}$, and in particular, the component oscillating at frequency $\omega$ reduces to
\begin{align}
\pb(t)={\rm Tr} \left\{ \db_{12}(\sigma_{12}^\dagger +\sigma_{12})\rho \right\} = 2 \db_{12}\,{\rm Re}\{\rho_{12}\} = 2 \db_{12} \left( \xi \ee^{-\ii\omega t} + \xi^* \ee^{\ii\omega t} \right) = \pb\,\ee^{-\ii\omega t} + {\rm c.c.},
\label{dsrho}
\end{align}
from which we obtain
\begin{align}
\pb = 2\db_{12}\,\xi = \frac{2\db_{12}\omega_0\eta}{(\omega + \ii\gamma_{21}/2)^2-\omega_0^2} \left(\rho_{22}-\rho_{11}\right).
\label{pfirst}
\end{align}

In order to derive the response of the atomic ensemble to the probe field, we consider the steady-state amplitude $\alpha_n$ of each photon mode of electric field $\eb_{nl}=\eb_n(\rb_l)$, evaluated at the position of every atom $l$. The coupling coefficients are then $g_{nl\,12} = -\db_{12}\cdot\eb_{nl}$. By using Eq.\ (\ref{dsrho}) in Eq.\ (\ref{dpcmeq}), we find
\begin{align}
\alpha_n = \beta^-_n\ee^{-\ii\omega t} + \beta^+ _n\ee^{\ii \omega t}
\nonumber
\end{align}
with coefficients
\begin{align}
\beta_n^- & = \frac{1}{\hbar(\omega_n-\omega-\ii\Gamma_n/2)}\sum_l\eb^*_{nl}\cdot\pb_l  , \nonumber\\
\beta_n^+  & = \frac{1}{\hbar(\omega_n+\omega-\ii\Gamma_n/2)}\sum_l\eb^*_{nl}\cdot\pb_l^*, \nonumber
\end{align}
where we reinsert the $l$ dependence in the sums over atoms. From here, using Eq.\ (\ref{gliibis}), we find
\begin{align}
g_{l12}+g_{l12}^* = \sum_n ( g_{nl12}\alpha_n + g_{nl12}^*\alpha_n^*) = -\db_{12}\cdot {\mathcal G}_{ll'}\cdot \pb_{l'}\ee^{-\ii\omega t} + {\rm c.c.},
\label{Eindgg}
\end{align}
where we define
\begin{align}
{\mathcal G}_{ll'} = \frac{1}{\hbar} \sum_n \left[ \frac{\eb_{nl}\otimes\eb_{nl'}^*}{\omega_n-\omega-\ii\Gamma_n/2} + \frac{\eb_{nl}^*\otimes\eb_{nl'}}{\omega_n+\omega+\ii\Gamma_n/2} \right]
\label{Gll}
\end{align}
as the electromagnetic Green tensor. As noted above [see Eq.\ (\ref{etaeta})], the total local field at frequency $\omega$ acting on emitter $l$ is given by the sum of the external field and the induced field produced by the emitters, that is, $\Eb_l^{\rm loc} = \Eb_l^{\rm probe} + \Eb_l^{\rm ind}$, where the induced field can be expressed, according to Eq.\ (\ref{Eindgg}), in terms of the Green tensor as $\Eb_l^{\rm ind}(t) = \sum_{l'} {\mathcal G}_{ll'}\cdot\pb_{l'}\ee^{-\ii\omega t} + {\rm c.c.}$, so the 
total local field becomes
\begin{align}
\Eb_l^{\rm loc}(t)=\left(\Eb_l^{\rm probe}+\sum_{l'} {\mathcal G}_{ll'}\cdot\pb_{l'}\right)\,\ee^{-\ii\omega t}+{\rm c.c.}
\nonumber
\end{align}
Finally, recalling that $(1/\hbar)\db_{12}\cdot\Eb_l^{\rm loc}(t) = \eta_l\ee^{-\ii\omega t}+{\rm c.c.}$ [see Eq.\ (\ref{etaeta})] and using Eq.\ (\ref{pfirst}), we obtain the coupled-dipole equations
\begin{align}
\pb_{l} = \tilde\alpha(\omega) \left[ \Eb_{l}^{\rm probe} + \sum_{l'}{\mathcal G}_{ll'}\cdot\pb_{l'} \right],
\label{CDEbis}
\end{align}
where
\begin{align}
\tilde\alpha(\omega) = \frac{2\omega_0\delta n}{\hbar}\frac{\db_{12}\otimes\db_{12}}{(\omega  + \ii \gamma_{21}/2 )^2 - \omega_0^2}
\nonumber
\end{align}
is the electrostatic polarizability tensor at a frequency $\omega$ near $\omega_0=\varepsilon_2-\varepsilon_1$, and [see Eqs.\ (\ref{finalrho})]
\begin{align}
\delta n = \rho_{22}-\rho_{11} = \frac{\displaystyle \left(\gamma_{21}^{-1}-\gamma_{32}^{-1}\right)\Delta'-1}{\displaystyle  1 + 2\gamma_{21}^{-1}\Delta + \left(\gamma_{21}^{-1}+2\gamma_{32}^{-1}\right)\Delta'+ 3\gamma_{21}^{-1}\gamma_{32}^{-1}\Delta\Delta'}
\label{Dn1}
\end{align}
is the population difference of the emitter.

Now, it is useful to recast Eq.\ (\ref{Dn1}) in terms of pump and local field amplitudes. We first rewrite Eq.\ (\ref{Delta}) using Eq.\ (\ref{etaeta}) as
\begin{align}
\Delta = \mathcal{I}\,\frac{\displaystyle 2\gamma_{21}^3\left( \omega^2 + \omega_0^2 + \gamma_{21}^2/4 \right)}{\left(\omega^2-\omega_0^2 - \gamma_{21}^2/4\right)^2 + \gamma_{21}^2\omega^2},
\nonumber
\end{align}
where we have assumed $\Eb^{\rm loc}$ to be oriented along $\db_{12}$, the $l$ dependence is again implicitly understood, and we have defined
\begin{align}
\mathcal{I} = \left|\Eb^{\rm loc}/E^{\rm probe}_{\rm sat}\right|^2,
\quad \quad
E^{\rm probe}_{\rm sat} = \hbar\gamma_{21}/d_{12}.
\nonumber
\end{align}
Additionally, we approximate $\gamma_{32}\ll\omega'$ and consider resonant pumping $\omega'=\omega'_0$, so that Eq.\ (\ref{Deltap}) together with Eq.\ (\ref{etapetap}) leads to
\begin{align}
\Delta' = \mathcal{I}'\,\gamma_{21},
\label{myDeltap}
\end{align}
where
\begin{align}
\mathcal{I}'=\left|\Eb^{\rm pump}/E^{\rm pump}_{\rm ref}\right|^2,
\quad \quad
E^{\rm pump}_{\rm ref} = (\hbar/2)\sqrt{\gamma_{21}\gamma_{32}}/d_{13}.
\label{Ippumpthres}
\end{align}
Incidentally, we note the relation $I=c|E|^2/2\pi$ between the light intensity $I$ and electric field $E$ in Gaussian units, which allows us to directly define a relation between $E^{\rm pump}$, $E^{\rm probe}$, and the corresponding pump and probe intensities.

We conclude by expressing some of the above quantities in terms of $\delta n$. In particular, using Eqs.\ (\ref{Dn1}) and  (\ref{myDeltap}), we find
\begin{align}
\Delta = \frac{1}{2}\;
\frac{(1-\mathcal{R})\mathcal{I}'-1
-\left[(1+2\mathcal{R})\mathcal{I}'+1\right]\,\delta n}{1+(3/2)\mathcal{R}\mathcal{I}'},
\label{deltadeltan}
\end{align}
which in turns allows us to rewrite Eqs.\ (\ref{eqrho11})-(\ref{eqrho33}) as
\begin{subequations}
\begin{align}
&\rho_{11}=\left[(1+\mathcal{R}\mathcal{I}')/(2+3\mathcal{R}\mathcal{I}')\right]\,(1-\delta n), \\
&\rho_{22}=\left[(1+\mathcal{R}\mathcal{I}')+
(1+2\mathcal{R}\mathcal{I}')\,\delta n\right]/(2+3\mathcal{R}\mathcal{I}'), \\
&\rho_{33}=\left[\mathcal{R}\mathcal{I}'/(2+3\mathcal{R}\mathcal{I}')\right]\,(1-\delta n),
\end{align}
\label{populationsfinal}
\end{subequations}
where
\begin{align}
\mathcal{R}=\gamma_{21}/\gamma_{32}.
\nonumber
\end{align}
Incidentally, in the $\gamma_{21}<<\gamma_{32}$ limit, assuming $\delta=\omega-\omega_0\ll\omega_0$ and $\gamma_{21}\ll\omega_0$, the population difference reduces to
\begin{align}
\delta n=\frac{-1+\mathcal{I}'}{1+8\mathcal{I}(1+3\mathcal{I}'\gamma_{21}/2\gamma_{32})/(1+4\delta^2/\gamma_{21}^2)+\mathcal{I}'}.
\nonumber
\end{align}

\subsection{Self-interaction and local density of optical states}
\label{radiativeone}

The self-interaction term $l'=l$ in Eq.\ (\ref{CDEbis}) involves $\mathcal{G}_{ll}$. The real part of this term exhibits a divergence originating in the electrostatic interaction of two point dipoles at vanishing separation (see also Sec.\ \ref{radiativetwo} below). This divergence can be effectively absorbed as an atomic resonance frequency shift. A detailed treatment of this effect requires a rather involved analysis \cite{FT02} that goes beyond the scope of the present work, so we just ignore it and assume it to be correctly incorporated in the atomic resonance frequency. However, the remaining imaginary part remains finite. In the limit of small mode decay rates $\Gamma_n$, we obtain from Eq.\ (\ref{Gll})
\begin{align}
{\rm Im}\left\{{\nt\cdot\mathcal G}_{ll}\cdot\nt\right\} =\frac{\pi}{\hbar} \sum_n \left|\nt\cdot\eb_{nl}\right|^2\,\delta(\omega_n-\omega)=2\pi^2\omega\,{\rm LDOS}_{\nt,l}(\omega),
\label{ImGLDOS}
\end{align}
where we define the frequency-dependent local density of optical states at the position of atom $l$ for polarization along a unit vector $\nt$ as
\begin{align}
{\rm LDOS}_{\nt,l}(\omega)=\frac{1}{2\pi\hbar\omega} \sum_n \left|\nt\cdot\eb_{nl}\right|^2\,\delta(\omega_n-\omega),
\nonumber
\end{align}
that is, the sum of all mode intensities. Note that the leading prefactor in this expression accounts for the normalization of the mode electric field as $\eb_{nl}/\sqrt{2\pi\hbar\omega}$, so that for example $|\eb_{nl}|^2=1/V$ in a free space of normalization volume $V$ (see Sec.\ \ref{Greentensorinfreespace} below).

Self-interaction can be understood as a radiative-reaction contribution to the response of the atom. Assuming an isotropic environment, one can conveniently absorb it in a corrected polarizability
\begin{align}
\alpha(\omega) = \frac{1}{1/\tilde\alpha(\omega)-2\pi^2\ii\omega\,{\rm LDOS}_{\nt,l}(\omega)},
\nonumber
\end{align}
while the coupled-dipole equations are modified as
\begin{align}
\pb_{l} = \alpha(\omega) \left[ \Eb_{l}^{\rm probe} + \sum_{l'\neq l}{\mathcal G}_{ll'}\cdot\pb_{l'} \right],
\label{CDE}
\end{align}
so that the $l'=l$ term is excluded from the sum.

\subsection{Atomic polarizability at the pump frequency}

By following a similar procedure, the electrostatic polarizability $\tilde\alpha(\omega')$ at the pump frequency $\omega'$ is obtained from the pump dipole
\begin{align}
\pb'(t)= 2 \db_{13} {\rm Re}\{\rho_{13}\} = 2 \db_{13} \left( \xi' \ee^{-\ii\omega't} + \xi'^* \ee^{\ii\omega't} \right) = \pb' \ee^{-\ii\omega't} + {\rm c.c.},
\nonumber
\end{align}
from which we find
\begin{align}
\tilde\alpha(\omega') = \frac{2\omega'_0\delta n'}{\hbar} \frac{\db_{13}\otimes\db_{13}}{(\omega'+\ii\gamma_{23}/2)^2-{\omega'}_0^2},
\label{alphawp}
\end{align}
where
\begin{align}
\delta n' = \rho_{33}-\rho_{11} = \frac{\displaystyle -\left(1 + \gamma_{21}^{-1}\Delta\right)}{\displaystyle  1 + 2\gamma_{21}^{-1}\Delta + \left(\gamma_{21}^{-1}+2\gamma_{32}^{-1}\right)\Delta'+ 3\gamma_{21}^{-1}\gamma_{32}^{-1}\Delta\Delta'}
\label{Dnp1}
\end{align}
This expression is valid for frequencies $\omega'$ near $\omega'_0$.

\section{Electromagnetic Green tensor, lattice sums, and reflectivity of 2D periodic arrays}

\subsection{Green tensor in free space}
\label{Greentensorinfreespace}

In this work, we focus on periodic planar atom arrays, with the atoms described through their polarizabilities and their interactions, self-consistently accounted for by means of Eq.\ (\ref{CDEbis}). For simplicity, we consider the arrays to be in vacuum. The electromagnetic Green tensor can then be worked out by using plane waves for the photon states,
$\eb_n(\rb_l) = -\ii\sqrt{2\pi\hbar qc/V}\,\ee^{\ii\qb\cdot \rb_l}\mbox{\boldmath$\hat{\bf \varepsilon}$}_{\sigma}$ with $\omega_n=qc$,
in which the mode index $n$ is multiplexed as $n \rightarrow\{\qb,\sigma\}$, where $\qb$ 
is the light wave vector, $\sigma=\,$s,\,p is the polarization state corresponding to the unit vector $\mbox{\boldmath$\hat{\bf \varepsilon}$}_{\sigma}$, and $V$ is the normalization volume. Additionally, the sum over photon modes becomes an integral using the substitution $\sum_n \rightarrow V\sum_{\sigma}\int d^3\qb/(2\pi)^3$. Also, we find the Green tensor $\mathcal{G}_{ll'}(\omega) = \mathcal{G}(\rb_{l}-\rb_{l'},\omega)$ to only depend on the relative coordinate vector $\rb=\rb_l-\rb_{l'}$. Putting these ingredients together, we find from Eq.\ (\ref{Gll})
\begin{align}
\mathcal{G}(\rb,\omega) = \int
\frac{d^3\qb}{(2\pi)^3}\frac{4\pi q^2\,\ee^{\ii\qb\cdot\rb}}{q^2-(k+i0^\dagger )^2}
\sum_{\sigma} \mbox{\boldmath$\hat{\bf \varepsilon}$}_{\sigma} \otimes
\mbox{\boldmath$\hat{\bf \varepsilon}$}_{\sigma},
\nonumber
\end{align}
where $k=\omega/c$ and we have replaced $\Gamma_n/2\rightarrow0^+$, as appropriate for photons in free space. Now, the sum over $\sigma$ can be transformed using the expression $\sum_{\sigma} \mbox{\boldmath$\hat{\bf \varepsilon}$}_{\sigma} \otimes
\mbox{\boldmath$\hat{\bf \varepsilon}$}_{\sigma}=\mathcal{I}_3-\hat{\qb}\otimes\hat{\qb}$, where $\mathcal{I}_3$ is the $3\times3$ identity matrix, while $\hat{\qb}$ is the unit vector along $\qb$. Additionally, $\qb$ can be replaced by $-\ii\nabla$. This leads to
\begin{align}
\mathcal{G}(\rb,\omega) &= \int \frac{d^3\qb}{(2\pi)^3} \,(\mathcal{I}_3q^2-\qb\otimes\qb)\,
\frac{4\pi\,\ee^{\ii\qb\cdot\rb}}{q^2-(k+i0^\dagger )^2} \nonumber\\
&= (-\nabla^2\mathcal{I}_3+\nabla\otimes\nabla) \int \frac{d^3\qb}{(2\pi)^3}\frac{4\pi\,\ee^{\ii\qb\cdot\rb}}{q^2-(k+i0^\dagger )^2} \nonumber \\
&= (k^2\mathcal{I}_3+\nabla\otimes\nabla)\frac{\ee^{\ii kr}}{r} \label{knana}  \\
&= \frac{\ee^{\ii kr}}{r^3}\left[(k^2r^2+\ii kr-1)\mathcal{I}_3-(k^2r^2+3\ii kr-3)\frac{\rb\otimes\rb}{r^2}\right], \nonumber 
\end{align}
which is the well-known dipole-dipole interaction tensor in free space. Incidentally, we have replaced $\nabla\rightarrow-k^2$ in the third line of the above derivation because $(\nabla^2+k^2)\ee^{\ii kr}/r=\delta(\rb)$ and we only need to evaluate the Green tensor for $\rb\neq0$.

\subsection{Radiative correction to the polarizability}
\label{radiativetwo}

As discussed in Sec.\ \ref{radiativeone}, the real part of $\mathcal{G}_{ll}$ diverges at short separations as the electrostatic dipole-dipole interaction $\nabla\otimes\nabla)(1/r)$, according to Eq.\ (\ref{knana}). The remaining imaginary part ${\rm Im}\{\mathcal{G}_{ll}\}=2\omega^3/3c^3$ is finite and produces a LDOS given by $\omega^2/3\pi^2c^3$ [see Eq.\ (\ref{ImGLDOS})], so that the corrected polarizability becomes
\begin{align}
\alpha(\omega) = \left\{\left[\frac{2\omega_0\delta n}{\hbar}\frac{\db_{12}\otimes\db_{12}}{(\omega  + \ii \gamma_{21}/2 )^2 - \omega_0^2}\right]^{-1}-\frac{2\ii\omega^3}{3c^3}\right\}^{-1}.
\label{alpharad}
\end{align}
Similarly, the polarizability at the pump frequency $\omega'$ can be written, including radiative corrections, as
\begin{align}
\alpha(\omega') = \left\{\left[\frac{2\omega'_0\delta n'}{\hbar} \frac{\db_{13}\otimes\db_{13}}{(\omega'+\ii\gamma_{23}/2)^2-{\omega'}_0^2}\right]^{-1}-\frac{2\ii\omega'^3}{3c^3}\right\}^{-1}.
\label{alphaprimerad}
\end{align}
We use Eqs.\ (\ref{alpharad}) and (\ref{alphaprimerad}) throughout this paper, so radiative corrections are incorporated in the polarizabilities, together with the coupled-dipole equations (\ref{CDE}) that exclude the $l'=l$ term.

\subsection{Lattice sums}

As we discuss in the main text, we consider a planar array formed by atoms sitting in the $z=0$ plane and having specular symmetry relative to the $x=0$ plane. For simplicity, we assume that the atoms can only be polarized along $y$ (i.e., all induced dipoles $\pb_l=p_l\,\yy$ are collinear and oriented along $y$). Under illumination by a plane wave, we then need to consider the incident electric-field component in the $z=0$ plane $E_y^{\rm probe}(x,y,0,t) = E_y^{\rm probe}\ee^{\ii\kparb\cdot\Rb-\ii\omega t}+{\rm c.c.}$, where $\omega$ is the frequency, $\kparb=(k_x,k_y)$ is the parallel component of the wave vector, and we use the notation $\Rb=(x,y)$.

Using Eq.\ (\ref{CDE}) and the methods described in more detail in Refs.\ \cite{paper090,paper182}, the induced dipole moment can be written as $p_l=p\,\ee^{\ii\kparb\cdot\Rb_l}$ with amplitude
\begin{align}
p = \frac{E^{\rm probe}}{1/\alpha(\omega)-G_{yy}(\kparb,\omega)},
\label{ppp}
\end{align}
where $G_{yy}(\kparb,\omega)$ is the $yy$ component of
\begin{align}
G(\kparb,\omega) = \sum_{l\not=0}\mathcal{G}(\Rb_l)\,\ee^{-\ii \kparb\cdot\Rb_l}, \label{LatticeSumOrigEq}
\end{align}
in which the sum runs over atomic lattice sites $\Rb_l$, omitting the atom at the origin $\Rb_{l=0}=0$. We now introduce the identities
\begin{subequations}
\begin{align}
& \frac{\ee^{\ii kr}}{r} = \ii\int\frac{d^2\Qb}{2\pi k_\perp^Q}\ee^{\ii\Qb\cdot\Rb+\ii k_\perp^Q|z|}, \label{identity1} \\
& \sum_l \ee^{\ii\Qb\cdot\Rb_l} = \frac{(2\pi)^2}{A}\sum_\gb\delta(\Qb-\gb),\label{identity2}
\end{align}
\label{identities}
\end{subequations}
where $\Qb=(Q_x,Q_y)$ is a 2D wave vector, $\gb$ runs over 2D reciprocal lattice vectors, $A$ is the unit-cell area,
\begin{align}
k_\perp^Q = \sqrt{k^2-Q^2+i0^+},
\nonumber
\end{align}
and the square root is taken to yield a positive imaginary part. Making use of Eqs.\ (\ref{knana}) and (\ref{identities}) for the evaluation of Eq.\ (\ref{LatticeSumOrigEq}), we find \cite{paper090,paper182}
\begin{align}
G_{yy}(\kparb,\omega) = \lim_{z\rightarrow 0}\left[\sum_\gb\frac{2\pi \ii}{Ak_\perp^{|\kparb+\gb|}}\exp\left(\ii k_\perp^{|\kparb+\gb|}|z|\right)\,\left[k^2-(k_y+g_y)^2\right]-\ii\int\frac{d^2\Qb}{2\pi k_\perp^Q}\ee^{\ii k_\perp^Q|z|}(k^2-Q_y^2)\right]. \label{latticesumeq}
\end{align}
Upon inspection of Eq.\ (\ref{latticesumeq}), we find that the imaginary part of $G_{yy}(\kparb,\omega)$ can be obtained analytically \cite{paper090,paper182} as
\begin{align}
{\rm Im}{\{G_{yy}(\kparb,\omega)\}}=\frac{2\pi}{A}\sum_\gb {\rm Re}\left\{\frac{\left[k^2-(k_y+g_y)^2\right]}{\sqrt{k^2-|\kparb+\gb|^2}}\right\}-2k^3/3,
\label{ImGyy}
\end{align}
while the remaining real part needs to be calculated numerically. The convergence of the series in Eq.\ (\ref{LatticeSumOrigEq}) is however slow, so we use the dedicated methods developed by Kambe \cite{K1968} in the context of low-energy electron diffraction.

\subsection{Specular reflectance}

The reflectance of the array can be now obtained by noticing that the electric field generated by an individual dipole $\pb$ placed at the origin is given by
\begin{align}
\Eb^{\rm dip}=[k^2\pb+(\pb\cdot\nabla)\nabla] \frac{\ee^{\ii kr}}{r}
\label{Edip}
\end{align}
[see Eq.\ (\ref{knana})]. Summing over all dipoles in the array and using Eqs.\ (\ref{identities}), the reflected field reduces to
\begin{align}
\Eb^{\rm ref} &=\ii p\,\sum_l\int\frac{d^2\Qb}{2\pi k^Q_\perp}\ee^{\ii\Qb\cdot(\Rb-\Rb_l)+\ii k^Q_\perp |z|}\ee^{\ii \kparb\cdot\Rb_l}\left[k^2\yy-Q_y\left(\Qb+{\rm sign}\{z\}k^Q_\perp\zz\right)\right]
\nonumber\\
&=\ii p\,\sum_\gb \Sb_\gb \,\exp\left[\ii(\kparb+\gb)\cdot\Rb+\ii k_\perp^{|\kparb+\gb|} |z|\right],
\nonumber
\end{align}
where
\begin{align}
\Sb_\gb=\frac{2\pi}{A k_\perp^{|\kparb+\gb|}}\,\left[k^2\yy-(k_y+g_y)\left(\kparb+\gb+{\rm sign}\{z\}k_\perp^{|\kparb+\gb|}\zz\right)\right].
\nonumber
\end{align}
For specular reflection ($\gb=0$), we have
\begin{align}
\Sb_0=\frac{2\pi}{A k_\perp^\kpar}\,\left[k^2\yy-k_y(\kparb+{\rm sign}\{z\}k_\perp^\kpar\zz)\right].
\nonumber
\end{align}
Because the dipoles are all oriented along $y$ regardless of the orientation of the incidence field, the array will reflect cross-polarized beams in general, unless $\kparb$ is directed along a symmetry direction of the array. For in/out s-polarization, the reflection coefficient reduces to
\begin{align}
r =\frac{\ii S}{1/\alpha(\omega)-G_{yy}(\kparb,\omega)},
\nonumber
\end{align}
where $S=2\pi k^2/Ak_\perp^\kpar$, which is the component considered in the main paper.

\begin{figure}[t]
\includegraphics[width=0.45\textwidth]{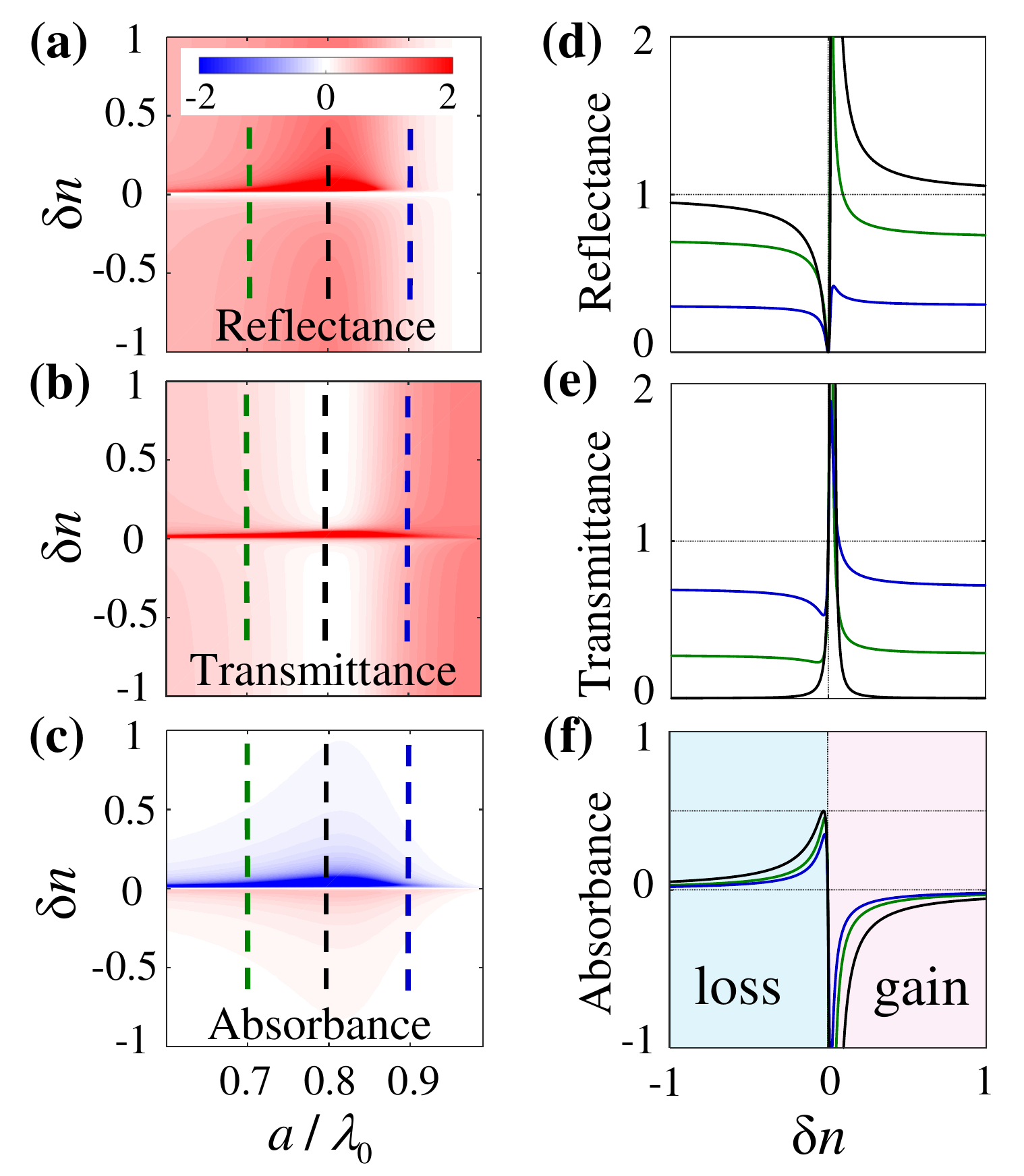}
\caption{{\bf (a-c)} Reflectance, transmittance, and absorbance under normal-incidence resonant-wavelength ($\lambda=\lambda_0=2\pi c/\omega_0$) probe illumination conditions as a function of population difference $\delta n$ and period-to-wavelength ratio $a/\lambda_0$. {\bf (d-f)} Cuts through (a-c) along the indicated vertical dashed lines with the same color code. We take the ratio between nonradiative and radiative scatterer decay rates to be $\gamma_{21}/\gamma_0=0.01$ in all plots. This figure is an extension of the data of Fig.\ 2 in the main paper.}
\label{FigS2}
\end{figure}

We plot in Fig.\ \ref{FigS2} an extension of Fig.\ 2 in the main paper in which we include reflectance and transmittance calculations as well.

\section{Continuous-wave lasing from a 2D atom array}

For simplicity, we assume transition dipole moments $\db_{12}$ and $\db_{13}$ oriented along $y$, so the interaction among induced dipoles in the array is described in terms of the $G_{yy}(\kparb,\omega)$ lattice sum alone. We also consider the pumping field at frequency $\omega'=\omega'_0$ to be along $y$. Laser emission takes place when the atoms acquire a polarization at frequency $\omega$ in the absence of an external probe field (i.e., $E^{\rm probe}=0$). Such a condition is signaled by the zeros in the denominator of Eq.\ (\ref{ppp}), that is,
\begin{align}
1/\alpha(\omega) = G_{yy}(\kparb,\omega).
\label{polecondition}
\end{align}
This clearly implies that lasing in the system is directional, as it occurs at specific directions $\kb_{\parallel}$ dictated by the noted condition.

\subsection{Lasing frequency and threshold pumping}

The emission frequency $\omega$ and pump intensity $I^{\rm pump}$ needed to have lasing are determined from Eq.\ (\ref{polecondition}). For convenience, we first subtract the term $-3\ii\omega^3/3c^3$ from both $1/\alpha(\omega)$ and $G_{yy}(\kparb,\omega)$ [see Eqs.\ (\ref{alpharad}) and (\ref{ImGyy})] and define
\begin{align}
&\mathfrak{g}_{\rm r}(\kparb,\omega)={\rm Re}\left\{G_{yy}(\kparb,\omega)\right\}, \nonumber \\
&\mathfrak{g}_{\rm i}(\kparb,\omega)={\rm Im}\left\{G_{yy}(\kparb,\omega)\right\}+3\omega^3/3c^3.
\nonumber
\end{align}
Using Eq.\ (\ref{latticesumeq}), we can write for $\mathfrak{g}_{\rm i}$ the explicit expression
\begin{align}
\mathfrak{g}_{\rm i}(\kparb,\omega) = \frac{2\pi}{A}\,\sum_\gb\left[k^2-(k_y+g_y)^2\right]\,{\rm Re}\left\{\frac{1}{k_\perp^{|\kparb+\gb|}}\right\}, \label{scaledG3}
\end{align}
which is in fact a finite sum contributed only by real diffracted components. Substituting Eq.\ (\ref{alpharad}) into the complex-number Eq.\ (\ref{polecondition}), we readily find it to be equivalent to the two real equations
\begin{subequations}
\begin{align}
&\omega^2-\omega_0^2-\gamma_{21}^2/4=\gamma_{21}\omega\frac{\mathfrak{g}_{\rm r}}{\mathfrak{g}_{\rm i}}, \label{wkpar} \\
&\delta n = \frac{1}{\alpha_0\mathfrak{g}_{\rm i}}, \label{deltanfinal}
\end{align}
\label{poleconditions}
\end{subequations}
where
\begin{align}
\alpha_0=2\omega_0 d_{12}^2/\hbar\gamma_{21}\omega.
\label{alpha0}
\end{align}
Equation\ (\ref{wkpar}) implicitly defines a resonance energy surface ($\omega$ as a function of $\kparb$), while Eq.\ (\ref{deltanfinal}) gives the induced dipole at the emission frequency $\omega$ needed to exactly compensate gains and losses. The latter requires a specific value of the induced dipole amplitude $p$, which we now calculate from Eq.\ (\ref{pfirst}) in combination with Eqs.\ (\ref{Delta}) and (\ref{deltanfinal}) to find $|p|^2=(\hbar\omega_0\omega/\mathfrak{g}_{\rm i})\Delta\delta n/(\omega^2+\omega_0^2+\gamma_{21}^2/4)$. Then, using Eq.\ (\ref{deltadeltan}) for $\Delta$, this expression reduces to
\begin{align}
|p|^2=\frac{\hbar\gamma_{21}}{2\mathfrak{g}_{\rm i}}\;
\frac{\omega_0\omega}{\omega^2 + \omega_0^2 + \gamma_{21}^2/4}\;
\frac{(1-\mathcal{R})\mathcal{I}'-1
-\left[(1+2\mathcal{R})\mathcal{I}'+1\right]\,\delta n}{1+(3/2)\mathcal{R}\mathcal{I}'},
\label{ppself}
\end{align}
where we recall that $\mathcal{R}=\gamma_{21}/\gamma_{32}$ and $\mathcal{I}'=I^{\rm pump}/I^{\rm pump}_{\rm ref}$ is the pump intensity normalized to a reference value defined as $I^{\rm pump}_{\rm ref}=c\hbar^2\gamma_{21}\gamma_{32}/8\pi d_{13}^2$ according to Eq.\ (\ref{Ippumpthres}). Importantly, the right-hand side of Eq.\ (\ref{ppself}) must be positive, a condition that imposes a threshold pumping
\begin{align}
\mathcal{I}'>\mathcal{I}'_{\rm thres}=\frac{1+\delta n}{1-\mathcal{R}-(1+2\mathcal{R})\,\delta n}
\label{Ipthreshold}
\end{align}
needed to sustain lasing. The transition dipole $d_{12}$ appears through $\alpha_0$ in these expressions [see Eq.\ (\ref{alpha0})], and obviously, it must reach a minimum value in order to enable lasing, as determined from the condition $\delta n=1/\alpha_0\mathfrak{g}_{\rm i}<1$, and also from the more restrictive condition imposed by the fact that the denominator in Eq.\ (\ref{Ipthreshold}) must be positive, leading to
\begin{align}
\alpha_0\mathfrak{g}_{\rm i}>\frac{1+2\mathcal{R}}{1-\mathcal{R}}.
\label{conditionag}
\end{align}

\subsection{Lasing intensity}

For the considered array of identical collinear dipoles, the laser intensity can be obtained from the far-field electric field, which is made up of contributions coming from all dipoles. Using Eqs.\ (\ref{identities}) and (\ref{Edip}), and introducing the dipole dependence on atom position $\Rb_l$ as $\pb_l=p\yy\ee^{\kparb\cdot\Rb_l}$, we find
\begin{align}
\Eb^{\rm laser}&=\ii p\sum_l \left(k^2\yy+\nabla\partial_{y}\right)
\int\frac{d^2\Qb'}{2\pi k_\perp^{Q'}}\ee^{\ii\Qb'\cdot\Rb+\ii k_\perp^{Q'}|z|}\ee^{\ii\kparb\cdot\Rb_l}
\nonumber \\
&=\sum_\gb
\Eb^{\rm laser}_\gb \ee^{\ii\Qb\cdot\Rb+\ii k_\perp^Q|z|},
\label{Egb}
\end{align}
where
\begin{align}
\Eb^{\rm laser}_\gb=
\frac{2\pi\ii p}{A k_\perp^Q}\left[k^2\yy-Q_y\left(Q_x,Q_y,k_\perp^Q{\rm sign}(z)\right)\right],
\label{Elaserg}
\end{align}
$\Qb = \kb_{\parallel} + \gb$, and $k^Q_\perp = \sqrt{k^2 - Q^2 + \ii0^+}$. Each reciprocal lattice vector $\gb$ in the sum of Eq.\ (\ref{Egb}) contributes with an emission intensity
\begin{align}
I^{\rm laser}_\gb=2\times \frac{c\left|\Eb^{\rm laser}_\gb\right|^2}{2\pi}\times {\rm Re}\left\{k^Q_\perp/k\right\},
\label{Ilaserg}
\end{align}
where the leading factor of 2 originates in the fact that each in-plane wave vector $\kparb$ involves identical emission toward both positive and negative $z$ directions, while the factor ${\rm Re}\left\{k^Q_\perp/k\right\}$ selects non-evanescent beams and compensates for the projection of the propagation direction on the normal direction $z$. Combining Eqs.\ (\ref{Elaserg}) and (\ref{Ilaserg}), comparing the result with Eq.\ (\ref{scaledG3}), and using Eq.\ (\ref{ppself}), we readily find
\begin{align}
I^{\rm laser} = \frac{\hbar\omega_0\gamma_{21}}{A}\,
\frac{\omega^2}{\omega^2 + \omega_0^2 + \gamma_{21}^2/4}\;
\frac{(1-\mathcal{R})\mathcal{I}'-1
-\left[(1+2\mathcal{R})\mathcal{I}'+1\right]\,\delta n}{1+(3/2)\mathcal{R}\mathcal{I}'}
\label{Ilaser}
\end{align} 
for the total laser emission intensity, where the unit-cell area $A$ emerges as a natural normalization.

\subsection{Energy conservation}

The energy stored in the system per unit area is $\mathcal{Q}=\hbar(\omega_0\rho_{22}+\omega'_0\rho_{33})/A$. Under steady-state conditions, energy conservation requires that the net power balance $\dot{\mathcal{Q}}$ between pump, lasing, and absorption channels should vanish:
\begin{align}
\dot{\mathcal{Q}} = I^{\rm pump} \mathcal{A}' - I^{\rm laser} - I^{\rm nr} = 0, \label{EnergyConsEquno}
\end{align}
where [see Eq.\ (\ref{Dnp1})]
\begin{align}
\mathcal{A}'\approx(4\pi\omega'_0/c){\rm Im}\{\alpha(\omega'_0)\}/A=8\pi\omega'_0d_{13}^2 (\rho_{11}-\rho_{33})/\hbar c \gamma_{32}A
\nonumber
\end{align}
is the absorbance at the pump frequency $\omega'_0$ [we use Eq.\ (\ref{alphawp}) and neglect lattice effects to obtain this expression, assuming that the pump frequency $\omega'_0$ is far from the lattice resonances and that the optical cross-section at that frequency is strongly reduced due to the large nonradiative damping $\gamma_{32}$], $I^{\rm pump}$ is the pump intensity,
\begin{align}
I^{\rm nr} = \rho_{22}\gamma_{21}\hbar\omega_0/A
+\rho_{33}\gamma_{32}\hbar(\omega'_0-\omega_0)/A
\nonumber
\end{align}
is the intensity dissipated by the system through nonradiative decay at rates $\gamma_{21}$ from level 2 and $\gamma_{32}$ from level 3, and $I^{\rm laser}$ is the lasing intensity [Eq.\ (\ref{Ilaser})].

We now evaluate all terms in Eq.\ (\ref{EnergyConsEquno}) using Eqs.\ (\ref{populationsfinal}), (\ref{deltanfinal}), and (\ref{Ilaser}),  which lead to the simple condition
\begin{align}
\omega=\sqrt{\omega_0^2+\gamma_{21}^2/4}
\label{wforlasing}
\end{align}
for energy conservation. Remarkably, this condition is independent of both $\kparb$ and $I^{\rm pump}$. For this value of the emission frequency, the lasing intensity further simplifies to
\begin{align}
I^{\rm laser} = \frac{\hbar\omega_0\gamma_{21}}{2A}\;
\frac{(1-\mathcal{R})\mathcal{I}'-1
-\left[(1+2\mathcal{R})\mathcal{I}'+1\right]\,\delta n}{1+(3/2)\mathcal{R}\mathcal{I}'},
\label{Ilaserfinal}
\end{align}
whereas the condition (\ref{wkpar}) reduces to $\mathfrak{g}_{\rm r}={\rm Re}\{G_{yy}\}=0$.

\subsection{Lasing stability}

For fixed lattice period, pump intensity, and atom characteristics, lasing can occur for different values of the parallel wave vector $\kparb$, subject to the conditions for stability
\begin{subequations}
\begin{align}
&(\partial\dot{\mathcal{Q}}/\partial k_x)(\partial\mathcal{Q}/\partial k_x)<0, \nonumber\\
&(\partial\dot{\mathcal{Q}}/\partial k_y)(\partial\mathcal{Q}/\partial k_y)<0. \nonumber
\end{align} 
\end{subequations}
This means that if $\kparb$ fluctuates away from equilibrium in a way such that $\mathcal{Q}$ becomes too large or two small then $\dot{\mathcal{Q}}$ decreases or increases in order to restore equilibrium.

\subsection{Calculation procedure}

For the results presented in the main text we consider a square-lattice array (primitive vectors along $x$ and $y$) of 3-level atoms characterized by the following parameters:
\begin{itemize}
\item $a$: lattice period
\item $\omega_0$: frequency difference between levels 2 and 1
\item $\omega'_0$: frequency difference between levels 3 and 1
\item $\gamma_0=4\omega_0^3d_{12}^2/3\hbar c^3$: radiative decay rate from level 2, expressed in terms of the $1\leftrightarrow2$ transition dipole $d_{12}$
\item $\gamma_{21}$ and $\gamma_{32}$: inelastic decay rates from levels 2 and 3, defining the ratio $\mathcal{R}=\gamma_{21}/\gamma_{32}\ll1$
\item $\mathcal{I}'=I^{\rm pump}/I^{\rm pump}_{\rm ref}$: normalized pump intensity at frequency $\omega'_0$, with $I^{\rm pump}_{\rm ref}=c\hbar^2\gamma_{21}\gamma_{32}/8\pi d_{13}^2$ expressed in terms of the $1\leftrightarrow3$ transition dipole $d_{13}$
\end{itemize}
The above parameters determine the emission frequency $\omega=\sqrt{\omega_0^2+\gamma_{21}^2/4}$ [Eq.\ (\ref{wforlasing})], subject to the conditions
\begin{align}
{\rm Re}\{G_{yy}(\kparb,\omega)\}=0
\nonumber
\end{align}
and
\begin{align}
\frac{1}{\delta n}=\frac{\omega_0}{\omega}\frac{\gamma_0}{\gamma_{21}}\frac{3\pi}{(\omega_0a/c)^3}\,f>\frac{1+2\mathcal{R}}{1-\mathcal{R}}
\nonumber
\end{align}
[see Eq.\ (\ref{conditionag})], where
\begin{align}
f=\sum_{m,n} {\rm Re}\left\{\frac{(ka)^2-(k_ya+2\pi n)^2}{\sqrt{(ka)^2-(k_xa+2\pi m)^2-(k_ya+2\pi n)^2}}\right\},
\nonumber
\end{align}
while $m$ and $n$ run over integer numbers labeling reciprocal lattice vectors $\gb=(2\pi/a)\,(m,n)$. The laser emission intensity reduces to
\begin{align}
I^{\rm laser} = \frac{\hbar\omega_0\gamma_{21}}{2a^2}\;
\frac{(1-\mathcal{R})\mathcal{I}'-1
-\left[(1+2\mathcal{R})\mathcal{I}'+1\right]\,\delta n}{1+(3/2)\mathcal{R}\mathcal{I}'}
\nonumber
\end{align}
[see Eq.\ (\ref{Ilaserfinal})].

\end{widetext}

\acknowledgments

This work has been supported in part by ERC (Advanced Grant 789104-eNANO), the Spanish MINECO (MAT2017-88492-R, SEV2015-0522, and PCIN-2015-155), AGAUR (2014 SGR 1400), and Fundaci\'o Privada Cellex.


\begin{thebibliography}{28}
\expandafter\ifx\csname natexlab\endcsname\relax\def\natexlab#1{#1}\fi
\expandafter\ifx\csname bibnamefont\endcsname\relax
  \def\bibnamefont#1{#1}\fi
\expandafter\ifx\csname bibfnamefont\endcsname\relax
  \def\bibfnamefont#1{#1}\fi
\expandafter\ifx\csname citenamefont\endcsname\relax
  \def\citenamefont#1{#1}\fi
\expandafter\ifx\csname url\endcsname\relax
  \def\url#1{\texttt{#1}}\fi
\expandafter\ifx\csname urlprefix\endcsname\relax\def\urlprefix{URL }\fi
\providecommand{\bibinfo}[2]{#2}
\providecommand{\eprint}[2][]{\url{#2}}

\bibitem[{\citenamefont{{Lord Rayleigh}}(1907)}]{R1907}
\bibinfo{author}{\bibnamefont{{Lord Rayleigh}}}, \bibinfo{journal}{Philos.\
  Mag.} \textbf{\bibinfo{volume}{14}}, \bibinfo{pages}{60}
  (\bibinfo{year}{1907}).

\bibitem[{\citenamefont{Wood}(1902)}]{W1902}
\bibinfo{author}{\bibfnamefont{R.~W.} \bibnamefont{Wood}},
  \bibinfo{journal}{Philos.\ Mag.} \textbf{\bibinfo{volume}{4}},
  \bibinfo{pages}{396} (\bibinfo{year}{1902}).

\bibitem[{\citenamefont{Fano}(1936)}]{F1936}
\bibinfo{author}{\bibfnamefont{U.}~\bibnamefont{Fano}},
  \bibinfo{journal}{Phys.\ Rev.} \textbf{\bibinfo{volume}{50}},
  \bibinfo{pages}{573} (\bibinfo{year}{1936}).

\bibitem[{\citenamefont{Ebbesen et~al.}(1998)\citenamefont{Ebbesen, Lezec,
  Ghaemi, Thio, and Wolff}}]{ELG98}
\bibinfo{author}{\bibfnamefont{T.~W.} \bibnamefont{Ebbesen}},
  \bibinfo{author}{\bibfnamefont{H.~J.} \bibnamefont{Lezec}},
  \bibinfo{author}{\bibfnamefont{H.~F.} \bibnamefont{Ghaemi}},
  \bibinfo{author}{\bibfnamefont{T.}~\bibnamefont{Thio}}, \bibnamefont{and}
  \bibinfo{author}{\bibfnamefont{P.~A.} \bibnamefont{Wolff}},
  \bibinfo{journal}{Nature} \textbf{\bibinfo{volume}{391}},
  \bibinfo{pages}{667} (\bibinfo{year}{1998}).

\bibitem[{\citenamefont{{Garc\'{\i}a de Abajo}}(2007)}]{paper090}
\bibinfo{author}{\bibfnamefont{F.~J.} \bibnamefont{{Garc\'{\i}a de Abajo}}},
  \bibinfo{journal}{Rev.\ Mod.\ Phys.} \textbf{\bibinfo{volume}{79}},
  \bibinfo{pages}{1267} (\bibinfo{year}{2007}).

\bibitem[{\citenamefont{{Garc\'{\i}a de Abajo}
  et~al.}(2005)\citenamefont{{Garc\'{\i}a de Abajo}, {G\'{o}mez-Medina}, and
  S\'{a}enz}}]{paper064}
\bibinfo{author}{\bibfnamefont{F.~J.} \bibnamefont{{Garc\'{\i}a de Abajo}}},
  \bibinfo{author}{\bibfnamefont{R.}~\bibnamefont{{G\'{o}mez-Medina}}},
  \bibnamefont{and} \bibinfo{author}{\bibfnamefont{J.~J.}
  \bibnamefont{S\'{a}enz}}, \bibinfo{journal}{Phys.\ Rev.\ E}
  \textbf{\bibinfo{volume}{72}}, \bibinfo{pages}{016608}
  (\bibinfo{year}{2005}).

\bibitem[{\citenamefont{Augui\'e and Barnes}(2008)}]{AB08}
\bibinfo{author}{\bibfnamefont{B.}~\bibnamefont{Augui\'e}} \bibnamefont{and}
  \bibinfo{author}{\bibfnamefont{W.~L.} \bibnamefont{Barnes}},
  \bibinfo{journal}{Phys.\ Rev.\ Lett.} \textbf{\bibinfo{volume}{101}},
  \bibinfo{pages}{143902} (\bibinfo{year}{2008}).

\bibitem[{\citenamefont{Zumofen et~al.}(2008)\citenamefont{Zumofen, Mojarad,
  Sandoghdar, and Agio}}]{ZMS08}
\bibinfo{author}{\bibfnamefont{G.}~\bibnamefont{Zumofen}},
  \bibinfo{author}{\bibfnamefont{N.~M.} \bibnamefont{Mojarad}},
  \bibinfo{author}{\bibfnamefont{V.}~\bibnamefont{Sandoghdar}},
  \bibnamefont{and} \bibinfo{author}{\bibfnamefont{M.}~\bibnamefont{Agio}},
  \bibinfo{journal}{Phys.\ Rev.\ Lett.} \textbf{\bibinfo{volume}{101}},
  \bibinfo{pages}{180404} (\bibinfo{year}{2008}).

\bibitem[{\citenamefont{Rezus et~al.}(2012)\citenamefont{Rezus, Walt, Lettow,
  Renn, Zumofen, {G\"otzinger}, and Sandoghdar}}]{RWL12}
\bibinfo{author}{\bibfnamefont{Y.~L.~A.} \bibnamefont{Rezus}},
  \bibinfo{author}{\bibfnamefont{S.~G.} \bibnamefont{Walt}},
  \bibinfo{author}{\bibfnamefont{R.}~\bibnamefont{Lettow}},
  \bibinfo{author}{\bibfnamefont{A.}~\bibnamefont{Renn}},
  \bibinfo{author}{\bibfnamefont{G.}~\bibnamefont{Zumofen}},
  \bibinfo{author}{\bibfnamefont{S.}~\bibnamefont{{G\"otzinger}}},
  \bibnamefont{and}
  \bibinfo{author}{\bibfnamefont{V.}~\bibnamefont{Sandoghdar}},
  \bibinfo{journal}{Phys.\ Rev.\ Lett.} \textbf{\bibinfo{volume}{108}},
  \bibinfo{pages}{093601} (\bibinfo{year}{2012}).

\bibitem[{\citenamefont{Hood et~al.}(2016)\citenamefont{Hood, Goban,
  {Asenjo-Garcia}, Lu, Yu, Chang, and Kimble}}]{HGA16}
\bibinfo{author}{\bibfnamefont{J.~D.} \bibnamefont{Hood}},
  \bibinfo{author}{\bibfnamefont{A.}~\bibnamefont{Goban}},
  \bibinfo{author}{\bibfnamefont{A.}~\bibnamefont{{Asenjo-Garcia}}},
  \bibinfo{author}{\bibfnamefont{M.}~\bibnamefont{Lu}},
  \bibinfo{author}{\bibfnamefont{S.-P.} \bibnamefont{Yu}},
  \bibinfo{author}{\bibfnamefont{D.~E.} \bibnamefont{Chang}}, \bibnamefont{and}
  \bibinfo{author}{\bibfnamefont{H.~J.} \bibnamefont{Kimble}},
  \bibinfo{journal}{PNAS} \textbf{\bibinfo{volume}{113}},
  \bibinfo{pages}{10507} (\bibinfo{year}{2016}).

\bibitem[{\citenamefont{Shahmoon et~al.}(2017)\citenamefont{Shahmoon, Wild,
  Lukin, and Yelin}}]{SWL17}
\bibinfo{author}{\bibfnamefont{E.}~\bibnamefont{Shahmoon}},
  \bibinfo{author}{\bibfnamefont{D.~S.} \bibnamefont{Wild}},
  \bibinfo{author}{\bibfnamefont{M.~D.} \bibnamefont{Lukin}}, \bibnamefont{and}
  \bibinfo{author}{\bibfnamefont{S.~F.} \bibnamefont{Yelin}},
  \bibinfo{journal}{Phys.\ Rev.\ Lett.} \textbf{\bibinfo{volume}{118}},
  \bibinfo{pages}{113601} (\bibinfo{year}{2017}).

\bibitem[{\citenamefont{Imada et~al.}(1999)\citenamefont{Imada, Noda, Chutinan,
  Tokuda, Murata, and Sasaki}}]{INC99}
\bibinfo{author}{\bibfnamefont{M.}~\bibnamefont{Imada}},
  \bibinfo{author}{\bibfnamefont{S.}~\bibnamefont{Noda}},
  \bibinfo{author}{\bibfnamefont{A.}~\bibnamefont{Chutinan}},
  \bibinfo{author}{\bibfnamefont{T.}~\bibnamefont{Tokuda}},
  \bibinfo{author}{\bibfnamefont{M.}~\bibnamefont{Murata}}, \bibnamefont{and}
  \bibinfo{author}{\bibfnamefont{G.}~\bibnamefont{Sasaki}},
  \bibinfo{journal}{Appl.\ Phys.\ Lett.} \textbf{\bibinfo{volume}{75}},
  \bibinfo{pages}{316} (\bibinfo{year}{1999}).

\bibitem[{\citenamefont{Meier et~al.}(1999)\citenamefont{Meier, Mekis,
  Dodabalapur, Timko, Slusher, Joannopoulos, and Nalamasu}}]{MMD99}
\bibinfo{author}{\bibfnamefont{M.}~\bibnamefont{Meier}},
  \bibinfo{author}{\bibfnamefont{A.}~\bibnamefont{Mekis}},
  \bibinfo{author}{\bibfnamefont{A.}~\bibnamefont{Dodabalapur}},
  \bibinfo{author}{\bibfnamefont{A.}~\bibnamefont{Timko}},
  \bibinfo{author}{\bibfnamefont{R.}~\bibnamefont{Slusher}},
  \bibinfo{author}{\bibfnamefont{J.}~\bibnamefont{Joannopoulos}},
  \bibnamefont{and} \bibinfo{author}{\bibfnamefont{O.}~\bibnamefont{Nalamasu}},
  \bibinfo{journal}{Appl.\ Phys.\ Lett.} \textbf{\bibinfo{volume}{74}},
  \bibinfo{pages}{7} (\bibinfo{year}{1999}).

\bibitem[{\citenamefont{Noda et~al.}(2001)\citenamefont{Noda, Yokoyama, Imada,
  Chutinan, and Mochizuki}}]{NYI01}
\bibinfo{author}{\bibfnamefont{S.}~\bibnamefont{Noda}},
  \bibinfo{author}{\bibfnamefont{M.}~\bibnamefont{Yokoyama}},
  \bibinfo{author}{\bibfnamefont{M.}~\bibnamefont{Imada}},
  \bibinfo{author}{\bibfnamefont{A.}~\bibnamefont{Chutinan}}, \bibnamefont{and}
  \bibinfo{author}{\bibfnamefont{M.}~\bibnamefont{Mochizuki}},
  \bibinfo{journal}{Science} \textbf{\bibinfo{volume}{293}},
  \bibinfo{pages}{1123} (\bibinfo{year}{2001}).

\bibitem[{\citenamefont{Altug et~al.}(2006)\citenamefont{Altug, Englund, and
  Vu{\v{c}}kovi{\'c}}}]{AEV06}
\bibinfo{author}{\bibfnamefont{H.}~\bibnamefont{Altug}},
  \bibinfo{author}{\bibfnamefont{D.}~\bibnamefont{Englund}}, \bibnamefont{and}
  \bibinfo{author}{\bibfnamefont{J.}~\bibnamefont{Vu{\v{c}}kovi{\'c}}},
  \bibinfo{journal}{Nat.\ Phys.} \textbf{\bibinfo{volume}{2}},
  \bibinfo{pages}{484} (\bibinfo{year}{2006}).

\bibitem[{\citenamefont{Wu et~al.}(2015)\citenamefont{Wu, Buckley, Schaibley,
  Feng, Yan, Mandrus, Hatami, Yao, Vu{\v{c}}kovi{\'c}, Majumdar
  et~al.}}]{WBS15}
\bibinfo{author}{\bibfnamefont{S.}~\bibnamefont{Wu}},
  \bibinfo{author}{\bibfnamefont{S.}~\bibnamefont{Buckley}},
  \bibinfo{author}{\bibfnamefont{J.~R.} \bibnamefont{Schaibley}},
  \bibinfo{author}{\bibfnamefont{L.}~\bibnamefont{Feng}},
  \bibinfo{author}{\bibfnamefont{J.}~\bibnamefont{Yan}},
  \bibinfo{author}{\bibfnamefont{D.~G.} \bibnamefont{Mandrus}},
  \bibinfo{author}{\bibfnamefont{F.}~\bibnamefont{Hatami}},
  \bibinfo{author}{\bibfnamefont{W.}~\bibnamefont{Yao}},
  \bibinfo{author}{\bibfnamefont{J.}~\bibnamefont{Vu{\v{c}}kovi{\'c}}},
  \bibinfo{author}{\bibfnamefont{A.}~\bibnamefont{Majumdar}},
  \bibnamefont{et~al.}, \bibinfo{journal}{Nature}
  \textbf{\bibinfo{volume}{520}}, \bibinfo{pages}{69} (\bibinfo{year}{2015}).

\bibitem[{\citenamefont{Bergman and Stockman}(2003)}]{BS03}
\bibinfo{author}{\bibfnamefont{D.~J.} \bibnamefont{Bergman}} \bibnamefont{and}
  \bibinfo{author}{\bibfnamefont{M.~I.} \bibnamefont{Stockman}},
  \bibinfo{journal}{Phys.\ Rev.\ Lett.} \textbf{\bibinfo{volume}{90}},
  \bibinfo{pages}{027402} (\bibinfo{year}{2003}).

\bibitem[{\citenamefont{Kewes et~al.}(2017)\citenamefont{Kewes, Herrmann,
  Rodr\'{\i}guez-Oliveros, Kuhlicke, Benson, and Busch}}]{KHR17}
\bibinfo{author}{\bibfnamefont{G.}~\bibnamefont{Kewes}},
  \bibinfo{author}{\bibfnamefont{K.}~\bibnamefont{Herrmann}},
  \bibinfo{author}{\bibfnamefont{R.}~\bibnamefont{Rodr\'{\i}guez-Oliveros}},
  \bibinfo{author}{\bibfnamefont{A.}~\bibnamefont{Kuhlicke}},
  \bibinfo{author}{\bibfnamefont{O.}~\bibnamefont{Benson}}, \bibnamefont{and}
  \bibinfo{author}{\bibfnamefont{K.}~\bibnamefont{Busch}},
  \bibinfo{journal}{Phys.\ Rev.\ Lett.} \textbf{\bibinfo{volume}{118}},
  \bibinfo{pages}{237402} (\bibinfo{year}{2017}).

\bibitem[{\citenamefont{{van Beijnum} et~al.}(2013)\citenamefont{{van Beijnum},
  {van Veldhoven}, Geluk, {de Dood}, Gert, and {van Exter}}}]{VVG13}
\bibinfo{author}{\bibfnamefont{F.}~\bibnamefont{{van Beijnum}}},
  \bibinfo{author}{\bibfnamefont{P.~J.} \bibnamefont{{van Veldhoven}}},
  \bibinfo{author}{\bibfnamefont{E.~J.} \bibnamefont{Geluk}},
  \bibinfo{author}{\bibfnamefont{M.~J.~A.} \bibnamefont{{de Dood}}},
  \bibinfo{author}{\bibfnamefont{W.}~\bibnamefont{Gert}}, \bibnamefont{and}
  \bibinfo{author}{\bibfnamefont{M.~P.} \bibnamefont{{van Exter}}},
  \bibinfo{journal}{Phys.\ Rev.\ Lett.} \textbf{\bibinfo{volume}{110}},
  \bibinfo{pages}{206802} (\bibinfo{year}{2013}).

\bibitem[{\citenamefont{Yang et~al.}(2015)\citenamefont{Yang, Hoang, Dridi,
  Deeb, Mikkelsen, Schatz, and Odom}}]{YHD15}
\bibinfo{author}{\bibfnamefont{A.}~\bibnamefont{Yang}},
  \bibinfo{author}{\bibfnamefont{T.~B.} \bibnamefont{Hoang}},
  \bibinfo{author}{\bibfnamefont{M.}~\bibnamefont{Dridi}},
  \bibinfo{author}{\bibfnamefont{C.}~\bibnamefont{Deeb}},
  \bibinfo{author}{\bibfnamefont{M.~H.} \bibnamefont{Mikkelsen}},
  \bibinfo{author}{\bibfnamefont{G.~C.} \bibnamefont{Schatz}},
  \bibnamefont{and} \bibinfo{author}{\bibfnamefont{T.~W.} \bibnamefont{Odom}},
  \bibinfo{journal}{Nat.\ Commun.} \textbf{\bibinfo{volume}{6}}
  (\bibinfo{year}{2015}).

\bibitem[{\citenamefont{Scully and Zubairy}(1997)}]{SZ97}
\bibinfo{author}{\bibfnamefont{M.~O.} \bibnamefont{Scully}} \bibnamefont{and}
  \bibinfo{author}{\bibfnamefont{M.~S.} \bibnamefont{Zubairy}},
  \emph{\bibinfo{title}{Quantum optics}} (\bibinfo{publisher}{Cambridge
  University Press}, \bibinfo{address}{Cambridge}, \bibinfo{year}{1997}).

\bibitem[{\citenamefont{Stockman}(2010)}]{S10_2}
\bibinfo{author}{\bibfnamefont{M.~I.} \bibnamefont{Stockman}},
  \bibinfo{journal}{J.\ Opt.} \textbf{\bibinfo{volume}{12}},
  \bibinfo{pages}{024004} (\bibinfo{year}{2010}).

\bibitem[{arr({\natexlab{a}})}]{array2D_comment1}
\bibinfo{note}{We neglect interatomic interactions at the pump frequency
  because the corresponing atomic polarizability is small under the assumption
  of large $\gamma_{32}$.}

\bibitem[{\citenamefont{Thongrattanasiri
  et~al.}(2012)\citenamefont{Thongrattanasiri, Koppens, and {Garc\'{\i}a de
  Abajo}}}]{paper182}
\bibinfo{author}{\bibfnamefont{S.}~\bibnamefont{Thongrattanasiri}},
  \bibinfo{author}{\bibfnamefont{F.~H.~L.} \bibnamefont{Koppens}},
  \bibnamefont{and} \bibinfo{author}{\bibfnamefont{F.~J.}
  \bibnamefont{{Garc\'{\i}a de Abajo}}}, \bibinfo{journal}{Phys.\ Rev.\ Lett.}
  \textbf{\bibinfo{volume}{108}}, \bibinfo{pages}{047401}
  (\bibinfo{year}{2012}).

\bibitem[{\citenamefont{Ficek and Tanas}(2002)}]{FT02}
\bibinfo{author}{\bibfnamefont{Z.}~\bibnamefont{Ficek}} \bibnamefont{and}
  \bibinfo{author}{\bibfnamefont{R.}~\bibnamefont{Tanas}},
  \bibinfo{journal}{Phys.\ Rep.} \textbf{\bibinfo{volume}{372}},
  \bibinfo{pages}{369} (\bibinfo{year}{2002}).

\bibitem[{arr({\natexlab{b}})}]{array2D}
\bibinfo{note}{See supplementary material at
  http://link.aps.org/supplemental/xxx for more details of the theoretical
  formalism.}

\bibitem[{\citenamefont{Li et~al.}(2003)\citenamefont{Li, Stockman, and
  Bergman}}]{LSB03}
\bibinfo{author}{\bibfnamefont{K.~R.} \bibnamefont{Li}},
  \bibinfo{author}{\bibfnamefont{M.~I.} \bibnamefont{Stockman}},
  \bibnamefont{and} \bibinfo{author}{\bibfnamefont{D.~J.}
  \bibnamefont{Bergman}}, \bibinfo{journal}{Phys.\ Rev.\ Lett.}
  \textbf{\bibinfo{volume}{91}}, \bibinfo{pages}{227402}
  (\bibinfo{year}{2003}).

\bibitem[{\citenamefont{Kambe}(1968)}]{K1968}
\bibinfo{author}{\bibfnamefont{K.}~\bibnamefont{Kambe}}, \bibinfo{journal}{Z.\
  Naturforsch.\ A} \textbf{\bibinfo{volume}{23}}, \bibinfo{pages}{1280}
  (\bibinfo{year}{1968}).

\end{thebibliography}

\end{document}